\documentclass[namedreferences]{SolarPhysics}
\usepackage[optionalrh]{spr-sola-addons} 
\usepackage{graphicx}
\usepackage{color}
\usepackage{url}             

\newcommand{\adv}{    {\it Adv. Space Res.}}

\newcommand{\aap}{    {\it Astron. Astrophys.}}

\newcommand{\apj}{    {\it Astrophys. J.}}
\newcommand{\apjl}{   {\it Astrophys. J. Lett.}}
\newcommand{\apss}{   {\it Astrophys. Spa. Sci.}}

\newcommand{\jgr}{    {\it J. Geophys. Res.}}

\newcommand{\nat}{    {\it Nature}}

\newcommand{\pasj}{   {\it Publ. Astron. Soc. Japan}}

\newcommand{\solphys}{{\it Solar Phys.}}

\newcommand{\ssr}{    {\it Space Sci. Rev.}}

\begin{document}
\begin{article}
\begin{opening}

\title{Responsibility of a Filament Eruption for the Initiation of a Flare,
CME, and Blast Wave, and its Possible Transformation into a Bow
Shock}

\author{V.V.~\surname{Grechnev}$^{1}$\sep
        A.M.~\surname{Uralov}$^{1}$\sep
        I.V.~\surname{Kuzmenko}$^{2}$\sep
        A.A.~\surname{Kochanov}$^{1}$\sep
        I.M.~\surname{Chertok}$^{3}$\sep
        S.S.~\surname{Kalashnikov}$^{1}$}

\runningauthor{Grechnev et al.} \runningtitle{Filament eruptions,
flares, CMEs, and shock waves}

\institute{$^{1}$ Institute of Solar-Terrestrial Physics SB RAS,
                  Lermontov St.\ 126A, Irkutsk 664033, Russia
                  email: \url{grechnev@iszf.irk.ru} email: \url{uralov@iszf.irk.ru} email: \url{kochanov@iszf.irk.ru}\\
           $^{2}$ Ussuriysk Astrophysical Observatory, Solnechnaya
                  St. 21, Primorsky Krai, Gornotaezhnoe 692533, Russia
                  email: \url{kuzmenko_irina@mail.ru} \\
           $^{3}$ Pushkov Institute of Terrestrial Magnetism,
                  Ionosphere and Radio Wave Propagation (IZMIRAN), Troitsk, Moscow, 142190 Russia
                  email: \url{ichertok@izmiran.ru}}

\date{Received ; accepted }

\begin{abstract}

Multi-instrument observations of two filament eruptions on 24
February and 11 May 2011 suggest the following updated scenario
for eruptive flare, CME and shock wave evolution. An initial
destabilization of a filament results in stretching out of
magnetic threads belonging to its body and rooted in the
photosphere along the inversion line. Their reconnection leads to
i)~heating of parts of the filament or its environment,
ii)~initial development of the flare arcade cusp and ribbons, and
iii)~increasing similarity of the filament to a curved flux rope
and its acceleration. Then the pre-eruption arcade enveloping the
filament gets involved in reconnection according to the standard
model and continues to form the flare arcade and ribbons. The
poloidal magnetic flux in the curved rope developing from the
filament progressively increases and forces its toroidal
expansion. This flux rope impulsively expands and produces an MHD
disturbance, which rapidly steepens into a shock. The shock passes
through the arcade expanding above the filament and then freely
propagates ahead of the CME like a decelerating blast wave for
some time. If the CME is slow, then the shock eventually decays.
Otherwise, the frontal part of the shock changes into the
bow-shock regime. This was observed for the first time in the 24
February 2011 event. When reconnection ceases, the flux rope
relaxes and constitutes the CME core--cavity system. The expanding
arcade develops into the CME frontal structure. We also found that
reconnection in the current sheet of a remote streamer forced by
the shock's passage results in a running flare-like process within
the streamer responsible for a type II burst. The development of
dimming and various associated phenomena are discussed.
\end{abstract}

\end{opening}

\hspace{4.0 truecm}
\parbox{7.5 truecm}
{\textit{Dedicated to the memory of Mukul Kundu who inspired a
significant part of our study}}

\section{Introduction}
  \label{S-introduction}

\subsection{Challenges of Solar Eruptions}
  \label{S-challenges}

The causes of solar flares, their relations to filament eruptions,
coronal mass ejections (CMEs), and underlying processes have been
considered for several decades. A number of flare models have been
proposed. The `standard' flare model, referred to as `CSHKP' for
its contributors \cite{Car64,Sturrock66,Hirayama1974,Kopp76}, is
the most elaborated one, being extensively supported by
observations. In particular, \inlinecite{Hirayama1974} proposed
that the current sheet, in which the flare reconnection occurred,
formed due to the lift-off of a filament, whose eruption was
driven by a magnetohydrodynamic (MHD) instability of an increasing
current in the filament. This scenario directly associated a flare
with an eruption, and thus provided a basis for understanding the
initiation of CMEs. Later considerations adopted the scenario of
\inlinecite{Hirayama1974}, but the filaments (prominences) were
mainly assumed to be passive dense plasmas accumulated near the
bottom of eruptive flux ropes. This picture revealed from
non-flare-related eruptions of filaments outside of active regions
(ARs), is thought to also apply to the flare-related eruptions
from ARs. However, the causes of the eruptions themselves have not
been elaborated by the standard model; and the model did not
incorporate confined flares.

The idea of \inlinecite{Hirayama1974} was elaborated by
\citeauthor{Chen1989} (\citeyear{Chen1989,Chen1996}), who
developed a theory of the expansion of a magnetic flux rope. The
propelling force driving an eruption in the Chen's \textit{Flux
rope model} is the Lorentz force, which governs the torus
instability. The difficulty of this model is that the ongoing
injection of the poloidal magnetic flux is required to ensure the
expansion of a CME \cite{KrallChenSantoro2000}. The
\textit{Breakout} reconnection \cite{Antiochos1999,Lynch2008} can
help to overcome the magnetic tension of the transverse magnetic
flux above the rope.

Another model of eruptions, the \textit{Tether cutting model}
proposed by \inlinecite{Moore2001}, is currently popular. A case
study of an eruptive event lead
\inlinecite{SterlingMooreThompson2001} to a conclusion that the
`tether cutting' reconnection occurred after the eruption onset
only as a by-product, while this scenario might apply to different
events.

\inlinecite{InhesterBirnHesse1992} proposed the formation of the
poloidal magnetic flux due to reconnection in a sheared arcade,
based on the original idea of
\inlinecite{vanBallegooijenMartens1989} (\textit{cf.}
\citeauthor{Uralov1990a}, \citeyear{Uralov1990a, Uralov1990b}). This
scenario (see also \opencite{LongcopeBeveridge2007}) was confirmed
quantitatively in the comparison of the magnetic flux reconnected in
flares with the poloidal flux of the corresponding magnetic clouds
near Earth \cite{Qiu2007}. \inlinecite{Zhang2001} and
\citeauthor{Temmer2008} (\citeyear{Temmer2008,Temmer2010}) found a
temporal correspondence between the hard X-ray flare emission and
acceleration of a CME. This fact is considered as a further support
to the formation of the CME's poloidal flux in the course of
magnetic reconnection responsible for the corresponding flare.

The \textit{dual-filament model}
\cite{Uralov2002,Grechnev2006erupt} combines the processes
employed by the standard model with effects of joining two
filaments. The increased total twist in the combined filament
forces the development of the torus instability. The combination
of the backbone fields of the filaments creates the initial
propelling force. These processes induce the stretching the
numerous filament threads anchored in the photosphere and
reconnection between them, thus increasing the inner twist.
Reconnection in the enveloping arcade augments the outer twist, as
in the standard model. The formerly stable filament transforms
into a `mainspring'.

This brief overview shows that many years of observational and
theoretical studies of solar eruptions have not yet led to
consensus about their scenarios, relation to flares, and the
development of CMEs. It is possible that different scenarios
adequately describe the events of different types. Observational
limitations restricted the opportunities to verify the existing
concepts.

\subsection{Excitation of Shock Waves}
  \label{S-shock_waves}

Another subject of long-standing debates is related to shock waves
propagating in the corona. Their existence is evidenced by several
phenomena. Following \inlinecite{Uchida1968}, Moreton waves are
considered as lower skirts of coronal shock waves. Many `EUV waves'
are also candidates. Type II radio bursts and interplanetary shocks
provide further support. Nevertheless, the origin of shocks remains
controversial (see, \textit{e.g.}, \opencite{VrsnakCliver2008} for a
review).

The first historical concept ascribed the excitation of coronal
shock waves to the flare plasma pressure pulses. In the scenario
of \inlinecite{Hirayama1974}, `\textit{in front of the prominence
a shock wave may be generated and this might be the cause of
either type-II or moving type-IV burst}'. A popular scenario
favored by the \textit{in situ} observations of bow shocks ahead
of interplanetary CMEs (ICMEs) and studies of solar data
\cite{Cliver2004} relates the excitation of `CME-driven shocks' to
the outer surfaces of super-Alfv{\' e}nic CMEs.

\citeauthor{Grechnev2011_I} (\citeyear{Grechnev2011_I,
Grechnev2011_III, Grechnev2013_20061213, Grechnev2014_II}) argued
in favor of the shock wave excitation by impulsively erupting flux
ropes, similar to the conjecture of \inlinecite{Hirayama1974}. A
flux rope forming from the filament, initially located low in the
corona, sharply accelerates and produces an MHD disturbance. It
rapidly steepens into a shock above an AR, where the fast-mode
speed steeply falls off \cite{Afanasyev2013}, and then freely
propagates like a decelerating blast wave. One might expect that
such blast-wave-like shocks propagating ahead of CMEs either
eventually get transformed into bow shocks, if the CME is fast, or
decay into a weak disturbance otherwise.

\subsection{Data, Approaches, and Aims}
 \label{S-data}

The uncertainties in the scenarios of eruptions, CMEs, and
excitation of shocks are due to the limitations of observations.
Ground-based observations of eruptions carried out mostly in the
H$\alpha$ line are restricted by the loss of the opacity of
eruptions in their expansion, and the Doppler shift removing them
from the filter passband. Space-borne observations in the He\,{\sc
ii} 304~\AA\ line, which is well-suited for the detection of
prominences, were infrequent in the past. The loss of the LASCO/C1
coronagraph on SOHO in 1998 resulted in a large gap between the
observations of near-surface activity and white-light CMEs. SOHO/EIT
provided insufficient temporal sampling of eruptions and wave-like
disturbances. Multi-wavelength data of a high temporal and spatial
resolution from telescopes of modern space missions such as the
\textit{Sun Earth Connection Coronal and Heliospheric Investigation}
instrument suites (SECCHI; \opencite{Howard2008}) on the
twin-spacecraft \textit{Solar-Terrestrial Relations Observatory}
(STEREO; \opencite{Kaiser2008}), the \textit{Atmospheric Imaging
Assembly} (AIA) on the \textit{Solar Dynamics Observatory} (SDO;
\opencite{Lemen2012AIA}) crucially improve the situation.

An important contribution is provided by observations of radio
emission. It is generated by various mechanisms and carries
quantitative information about them. Microwave images show
filaments and prominences with a large field of view, which can
overlap with the images produced by the SOHO's \textit{Large Angle
and Spectroscopic Coronagraph} (LASCO; \opencite{Brueckner1995}).
Microwave images from the \textit{Nobeyama Radioheliograph} (NoRH;
\opencite{Nakajima1994}) and \textit{Siberian Solar Radio
Telescope} (SSRT; \opencite{Smolkov1986}; \opencite{Grechnev2003})
have been successfully used in studies of eruptions and related
phenomena (\textit{e.g.}, \opencite{Hanaoka1994};
\opencite{Uralov2002}; \opencite{Shimojo2006};
\citeauthor{Grechnev2006erupt}
\citeyear{Grechnev2006erupt,Grechnev2008shocks};
\opencite{Alissandrakis2013}). The dynamic radio spectra reveal
the signatures of propagating shocks (type IIs) and expanding
ejecta (type IVs).

Measurements of eruptive features, which are usually faint relative
to associated flares, are complicated by a rapid decrease of their
brightness or opacity. The difficulties to detect and follow them in
all analyzed images lead to large positional uncertainties. Even in
modern elaborations (\textit{e.g.}, \opencite{Vrsnak2007};
\opencite{Temmer2010}), the differentiation of the measured
distance-time points causes a large scatter of the velocities and
accelerations. This drawback is reduced in the approach based on the
fit of an analytic function to the measurements (\textit{e.g.},
\opencite{Gallagher2003}; \opencite{Sheeley2007};
\opencite{WangZhangShen2009}; \opencite{Alissandrakis2013}). The
approach uses the fact that the initial and final velocities of an
eruption are nearly constant, while the acceleration occurs within a
limited time interval. We fit the acceleration with a Gaussian time
profile. The kinematical plots are calculated by means of the
integration of the analytic fit rather than the differentiation of
the measurements. The results of the fit are used as a starting
estimate of the parameters of the acceleration, and then they are
optimized to outline the eruption in a best way. Our ultimate
criterion is to follow the analyzed feature as closely as possible
in all of the images. If the kinematics is more complex, then we use
a combination of Gaussians and adjust their parameters manually. The
major errors are due to the uncertainties in following a moving
feature, whose visibility decreases. All methods should converge to
similar results. The accuracy and performance of each method are
different. The accuracy is essential for our purposes.

Our measurements of wave signatures are based on the same approach
and a power-law fit expected for shock waves. The techniques are
described in \citeauthor{Grechnev2011_I} (\citeyear{Grechnev2011_I,
Grechnev2011_III, Grechnev2013_20061213, Grechnev2014_I,
Grechnev2014_II}).

Using multi-instrument observations in various spectral ranges, we
pursue a deeper insight into the following issues: how do
impulsive eruptions occur; how do CMEs form; when and where are
shock waves excited, and where they in the images are; are they
related to `EUV waves' or not. For these purposes we consider two
eruptive events of different energetics. We analyze the observed
relations between the filament eruptions and the onset of the
flares. To reveal the histories of the disturbances produced by
the eruptions, we detect and reconcile their manifestations at
different wavelengths such as the response of individual coronal
structures located not far from the eruption sites; large-scale
`EUV waves'; the outermost envelopes of the white-light CMEs; the
trajectories and structures of type II radio bursts.

Section~\ref{S-e1} addresses a strong event responsible for a
blast-wave-like shock, which then approached the bow-shock regime.
Section~\ref{S-e2} considers the development of a CME in a weak
event, also responsible for a blast-wave-like shock, which decayed
into a weak disturbance. The results are discussed in
Section~\ref{S-discussion} and summarized in
Section~\ref{S-conclusion}. The events and the measurements are
illustrated by the movies accompanying the electronic version of the
paper.

\section{Event I: 24 February 2011}
 \label{S-e1}

A prominence eruption associated with an M3.5 flare occurred on 24
February 2011 around 07:30 (\textit{all times hereafter refer to
UT}) on the east limb. The event was observed along the Sun--Earth
line by SDO/AIA, SOHO/LASCO, and the \textit{Reuven Ramaty
High-Energy Solar Spectroscopic Imager} (RHESSI;
\opencite{Lin2002}). The eruption site were also visible from
STEREO-B, which was located $94^{\circ}.55$ east of the Earth at a
distance of 1.022 AU from the Sun.

The eruption was analyzed by \inlinecite{Kumar2012}, who concluded
that the prominence accelerated due to the torus instability.
Considering in terms of high-beta conditions the role of the
plasma pressure in the initiation of the eruption, the authors
have not come to a certain conclusion because of the unknown
plasma density and magnetic field strength in the prominence,
which was supposed to be a part of a larger flux rope. We will
address the kinematics of the eruptive structures, their relation
to the flare, and follow the development of the shock wave and
CME.

\subsection{Filament Eruption}
 \label{S-e1_eruption}

\begin{figure} 
  \centerline{\includegraphics[width=\textwidth]
   {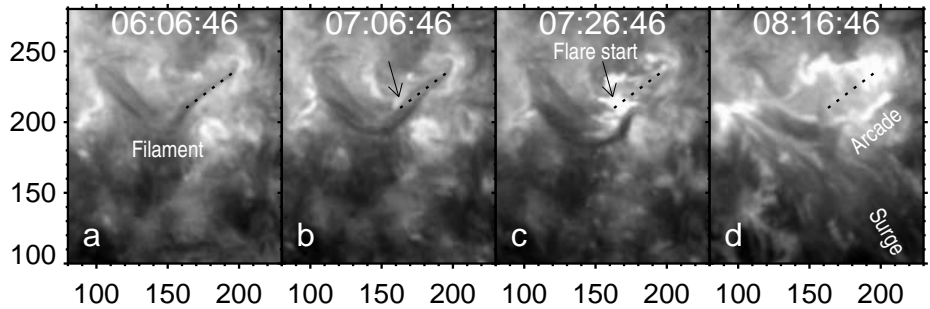}
  }
  \caption{The eruption region of the 24 February 2011 event observed
with STEREO-B/EUVI in 304~\AA: (a),~(b)~initiation, (c)~the onset
of the eruption and flare, d)~post-eruption arcade. The images
were rotated to 07:30. The black dotted bar marks the initial axis
of the west filament arm. The arrows in panels (b) and (c)
indicate the appearing flare brightening. The field of view
corresponds to the red frame in Figure~\ref{F-20110224_wave}a. The
axes show arc seconds from the solar disk center.}
  \label{F-20110224_initiation}
  \end{figure}

The initiation phase manifested in gradual changes of the V-shaped
filament visible in STEREO-B/EUVI 304~\AA\ images more than half
an hour before the eruption. In
Figure~\ref{F-20110224_initiation}a, the axis of the west filament
arm is traced with the dotted bar. No indications of a helical
structure are pronounced. In Figure~\ref{F-20110224_initiation}b,
the filament slightly straightened and broadened. A small region
indicated by the arrow started to brighten up. In
Figure~\ref{F-20110224_initiation}c, the filament is considerably
displaced. The flare brightening became conspicuous. Its
rhombus-like extensions were due to an overexposure distortion,
which contaminated the images during the flare.
Figure~\ref{F-20110224_initiation}d shows a post-eruption arcade
and a surge emanating from the east filament toward the southwest.
The axis of the west arcade portion was close to that of the
pre-eruption filament, in agreement with the CSHKP model.

\begin{figure} 
  \centerline{\includegraphics[width=0.7\textwidth]
   {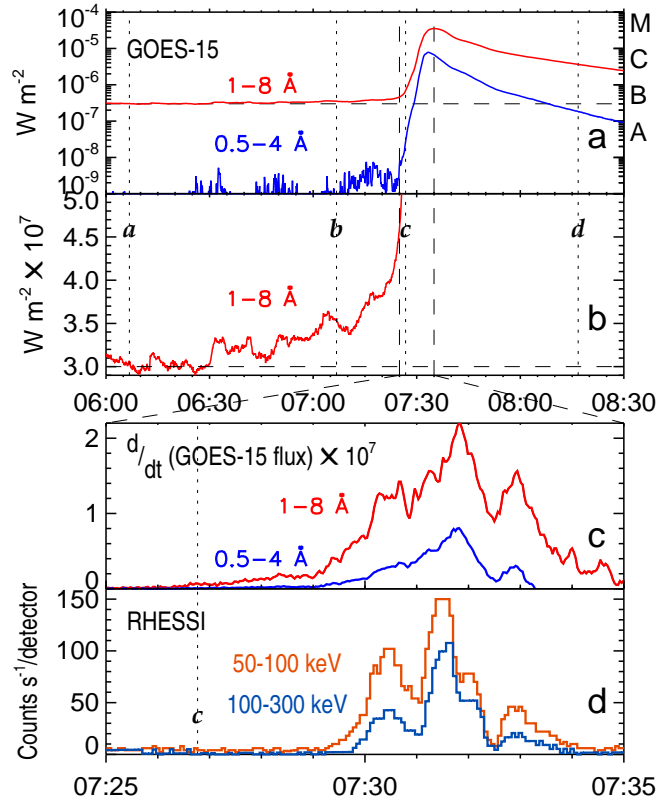}
  }
  \caption{GOES time profiles of the SXR emission in the
24 February 2011 event in the logarithmic (a) and linear (b)
scales, (c)~their derivatives, and (d)~hard X-ray burst recorded
with RHESSI. The dotted vertical lines mark the observation times
of the images in Figure~\ref{F-20110224_initiation}. The dashed
vertical lines in panels a) and b) mark the interval shown in
panels (c) and (d).}
  \label{F-20110224_rhessi_goes}
  \end{figure}

The activation of the filament was accompanied by a gradual rise of
the soft X-ray (SXR) flux in Figures \ref{F-20110224_rhessi_goes}a
and Figure~\ref{F-20110224_rhessi_goes}b. This observation supports
and elaborates the conjecture of \inlinecite{Zhang2001} about the
correspondence between the activation of a filament and a gradual
rise of the SXR flux.

The filament started to erupt at about 07:27 and separated (see
the movies 20110224\_euvi\_304\_fulldisk.mpg and
20110224\_euvi\_195\_fulldisk.mpg). Its fastest portion is
detectable in the 304~\AA\ image ratio at 08:06:46 above the
southwest limb at $-45^\circ$ from the west direction, along the
axis of AR~11164 (the arrow in Figure~\ref{F-20110224_wave}). We
adopt this angle as the initial orientation of the CME in the
plane of the sky of STEREO-B. This eruption moved perpendicularly
to a presumable arrangement of the streamer belt near this part of
the limb.

The flare site at a longitude of E84 was visible from the Earth.
Thus, RHESSI data provide complete information about hard X-ray
(HXR) emission of this flare
\cite{BattagliaKontar2012,MartinezOliveros2012}.
Figure~\ref{F-20110224_rhessi_goes}d presents the HXR burst in two
energy bands in comparison with the derivatives of two GOES
channels in Figure~\ref{F-20110224_rhessi_goes}c. The Neupert
effect \cite{Neupert1968} worked well, being somewhat energy
dependent. This fact confirms the simple scenario with
chromospheric evaporation caused by precipitating electrons and
confinement of evaporated plasmas in the flare loops emitting soft
X-rays \cite{Kumar2012}, unlike the events with more than one
eruption, when pronounced deviations from the Neupert effect can
occur (see, \textit{e.g.}, \opencite{Grechnev2013_20061213}).

\begin{figure} 
  \centerline{\includegraphics[width=\textwidth]
   {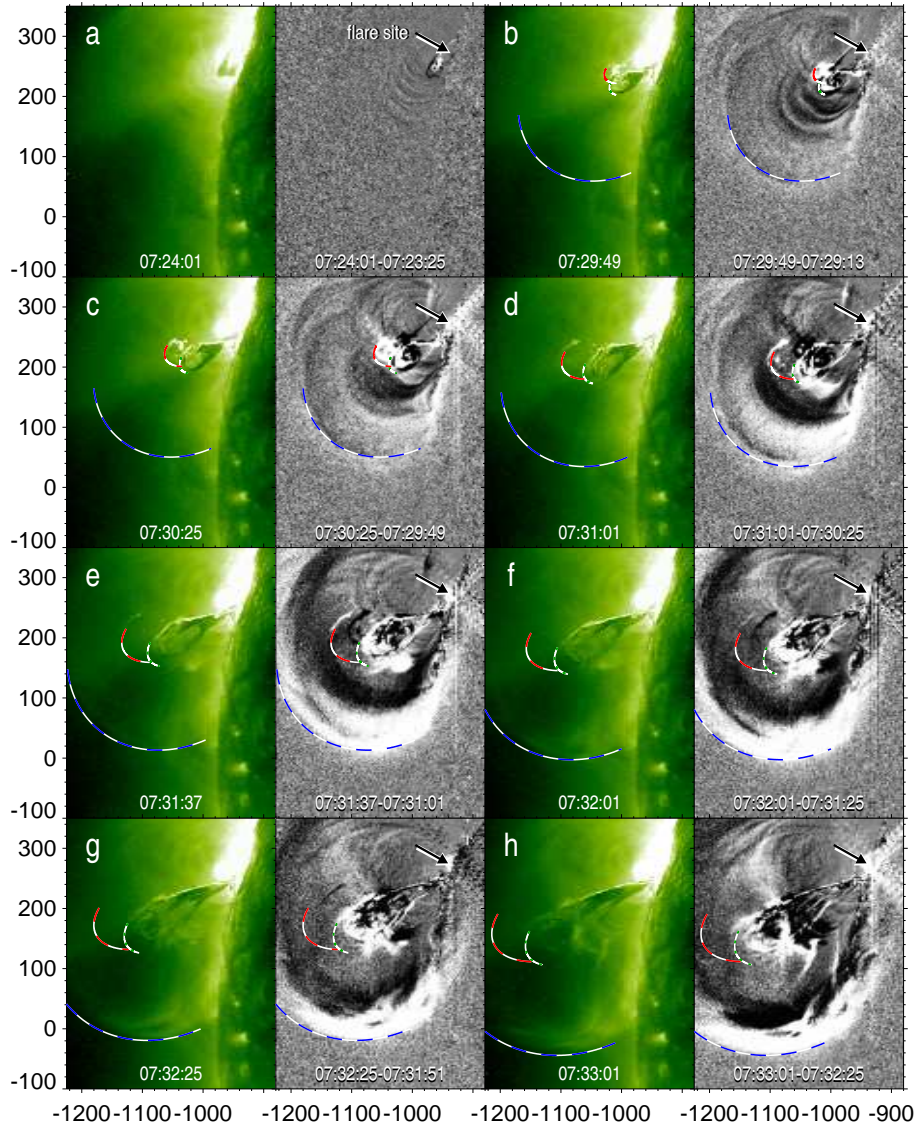}
  }
  \caption{Prominence eruption on 24 February 2011
observed by SDO/AIA in 211~\AA. The left, green panel of each pair
of the images from (a) to (h) presents an enhanced-contrast image,
and each right, gray-scale panel shows a running-difference ratio.
The dashed arcs outline the bright (red/white) and dark
(green/white) rings of the eruptive prominence and the overlying
arcade loop (blue/white) according to the kinematic measurements
in Figure~\ref{F-20110224_kinematics}. The arrow in the right
panels indicates the flare site. The axes show the coordinates in
arcsec from the solar disk center.}
  \label{F-20110224_sdo_images}
  \end{figure}

The detailed AIA data present a violent prominence eruption (see
also the movie 20110224\_AIA\_211\_eruption.mpg).
Figure~\ref{F-20110224_sdo_images} shows eight pairs of the AIA
211~\AA\ images (the characteristic temperature is 2~MK). The left
panel in each pair presents a 211~\AA\ image in the logarithmic
brightness scale; the right panel presents a running-difference
ratio of the left image to the pre-event one. The structure in the
figure looks like the cross section of a large flux rope, whose
axis had an acute angle with the line of sight, and passive
prominence material in its bottom part. However, the picture
revealed by the AIA data is more complex.

Initially, the prominence was dark and screened features behind it
such as a bright ring in Figure~\ref{F-20110224_sdo_images}c. This
fact indicates absorption, which occurs if the temperature of the
prominence body is low, $\lesssim 10^4$~K (see, \textit{e.g.},
\citeauthor{Grechnev2008shocks},
\citeyear{Grechnev2008shocks,Grechnev2014_I}). Then a helical
structure of the prominence appeared and evolved. The small ring
became bright at 211~\AA\ that indicates its heating up to $\gsim
2$~MK (or still higher, see \opencite{Kumar2012}). The brightening
of the formerly dark prominence suggests an increase of the plasma
pressure, $2nkT$, by a factor of $\gsim 200$ (\textit{cf}.
\opencite{Grechnev2014_I}). Such a strong pressure rise could
deform the prominence and inspire the development of some
instability. The rise of the SXR flux during the initiation phase
supports the role of heating.

\begin{figure} 
  \centerline{\includegraphics[width=0.7\textwidth]
   {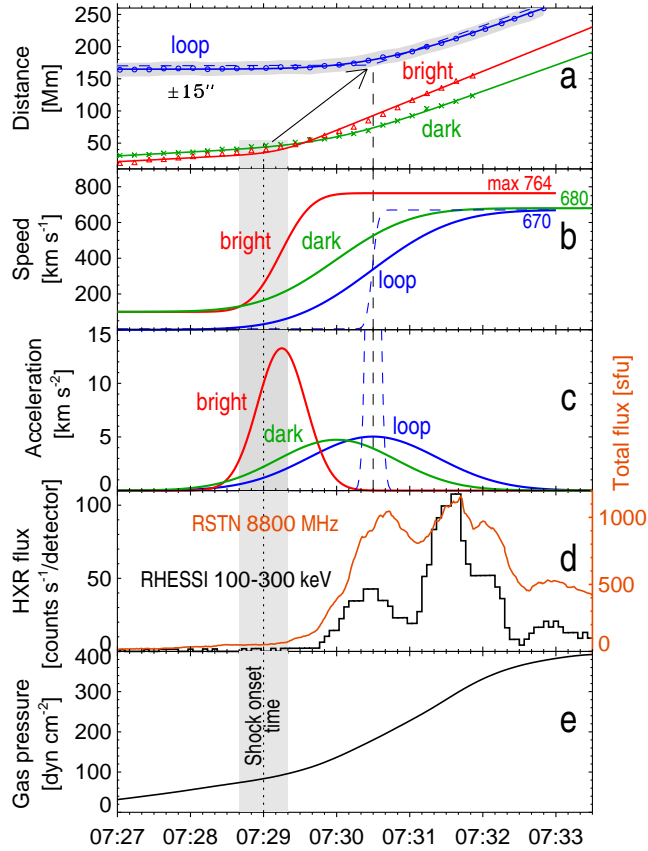}
  }
  \caption{Plane-of-the-sky kinematics of the eruptive
prominence and enveloping arcade measured from AIA images (a--c) in
comparison with HXR and microwave time profiles (d), and gas
pressure in flare loops (e). The symbols in panel (a) represent the
coarse starting measurements, and the gray error band corresponds to
$\pm 15^{\prime \prime}$. The blue dashed lines schematically show
the plots for a possible shorter acceleration of the arcade loop.
The dotted vertical line denotes the estimated shock onset time, and
the shaded interval presents its uncertainty. The dashed vertical
line corresponds to the acceleration center time of the arcade. The
arrow in panel (a) indicates a disturbance propagating from the
erupting prominence to the arcade. The tilt of the arrow corresponds
to the average speed of the disturbance of $\approx
1500$~km~s$^{-1}$ within the time interval between the ends of the
arrow.}
  \label{F-20110224_kinematics}
  \end{figure}

The gray-scale right panels of Figure~\ref{F-20110224_sdo_images}
show concentric loops of a pre-eruption arcade. The loops embraced
the prominence; their expansion lagged behind it. Figures
\ref{F-20110224_kinematics}a--\ref{F-20110224_kinematics}c present
the kinematics of two prominence segments and one of the outermost
loops, whose delay is most pronounced. The measurement technique
(\citeauthor{Grechnev2011_I} \citeyear{Grechnev2011_I,
Grechnev2014_I}) is briefly described in Section~\ref{S-data}. The
errors of $\pm 15^{\prime \prime}$ produce minor uncertainties in
the velocities of the filament segments of $\pm 4\%$ and
acceleration center times of $\pm 9$~s. The acceleration of the
arcade is more uncertain. Applying to the arcade loop a much
shorter, stronger acceleration with the same center time results
in an indiscernible change of the calculated positions. Therefore,
the solid blue plots for the loop in Figures
\ref{F-20110224_kinematics}a and \ref{F-20110224_kinematics}b are
the limits corresponding to the longest acceleration. The dashed
blue curves correspond to a shorter acceleration time. The
measurements are limited by the field of view of AIA; the final
speeds of the eruptive features can be higher.

The eruption in Figure~\ref{F-20110224_sdo_images} moved with an
angle of $\approx 25^{\circ}$ to the east direction visible from
SDO. Its orientation with respect to the west direction visible
from STEREO-B was $\approx 45^{\circ}$
(Figure~\ref{F-20110224_wave}). Using the nearly perpendicular
viewing direction of STEREO-B relative to the Sun--Earth line, one
can approximately estimate the velocity components directly from
the images observed with STEREO-B and AIA: the module of the
velocity and acceleration vectors should be related to the
plane-of-the-sky measurements from AIA data by a factor of about
1.09.

The bright helical ring of the eruptive prominence was a most
dynamic feature, leading all others. Its acceleration was the
highest one and reached $a_{1\,\mathrm{(POS)}} \approx
13$~km~s$^{-2}$ in the plane of the sky ($|\mathbf{a_1}| \approx
14$~km~s$^{-2} \approx 50g_{\odot}$; $g_{\odot} = 274$~m~s$^{-2}$
is the solar gravity acceleration at the photospheric level). The
darker ring accelerated somewhat later with $a_{2
\,\mathrm{(POS)}} \approx 5$~km~s$^{-2}$. The prominence drove the
expansion of the arcade, which lagged behind the fastest
prominence ring by about 1.5 min.

It is useful to compare the acceleration of the eruption with the
bursts in HXR (RHESSI) and microwaves (USAF RSTN San Vito and
Learmonth stations). The correspondence between HXR and microwave
bursts is well known. A temporal correspondence between HXR bursts
and CME acceleration was established by \citeauthor{Temmer2008}
(\citeyear{Temmer2008,Temmer2010}). Here we see that the latter
correspondence might be due to the delays of both the arcade
expansion and flare emission with respect to the acceleration of
the prominence, which itself could be the developing flux rope.
The prominence threads in Figure~\ref{F-20110224_sdo_images} were
connected to the flare site resembling a flare cusp. The dynamic
cusp formation was previously considered by
\inlinecite{SuiHolmanDennis2008}. The distance between the cusp
and the flare site was $\lesssim 50$~Mm. The delayed flare
development relative to the eruption agrees with the standard
model. The observations do not reveal any signs of the breakout
reconnection \cite{Antiochos1999} within the field of view of AIA;
the eruptive prominence apparently had to overcome the magnetic
tension of the overlying closed fields by itself. Neither the
images in Figure~\ref{F-20110224_sdo_images} show the lateral
overexpansion discussed by \inlinecite{Patsourakos2010}. The shape
of the arcade looks consistent with a limitation of its expansion
by the solar surface and a tilted motion of the eruptive
prominence.

Figure~\ref{F-20110224_kinematics}e presents the evolution of the
plasma pressure in flare loops computed from SXR GOES fluxes. The
temperature, $T$, and emission measure, EM, were calculated from
the SXR fluxes by means of the standard GOES software. The number
density, $n$, was estimated as $n = \sqrt{\mathrm{EM}/V}$, with a
volume $V \approx A^{3/2}$. To find the area, $A$, we used the
dimensions of the SXR-emitting source evaluated by
\inlinecite{BattagliaKontar2012} from RHESSI images. The pressure,
$2nkT$, steadily rose until a maximum at 07:33:44, and then
started to gradually decrease. The half-height duration of the
gradual pressure pulse was about 18~min. These characteristics of
the flare pressure rule out its significance in the initiation of
either the eruption or the wave, whose estimated onset time, $t_0
\approx $~07:29:00, is marked in
Figure~\ref{F-20110224_kinematics} with the dotted line. A similar
relation between the wave onset and flare pressure was shown for
different events by \inlinecite{Grechnev2011_I}. They also
presented the reasons why the flare-ignition of shocks is not
expected, in general.

The eruptive prominence underwent an impulsive acceleration up to
$|\mathbf{a}| \approx 50g_{\odot}$. This spurt produced a wave
disturbance with an onset time $t_0$. The disturbance traveled
about 150~Mm with an average plane-of-the-sky speed of $\approx
1500$~km~s$^{-1}$ (the arrow in
Figure~\ref{F-20110224_kinematics}a), at 07:30:30 arrived at the
arcade loop, which gradually expanded above the prominence, and
accelerated its lift-off.

With the active role of the prominence, whose eruption presented
presumable completion of the flux rope formation, the overlying
arcade expanded, being initially driven from inside. Generally, the
top part of an arcade is associated with a magnetic separatrix
surface, which does not allow plasmas to be transposed inside from
outside. The right panels of Figure~\ref{F-20110224_sdo_images} show
that the compression region of swept-up plasmas on top of the
expanding arcade evolved. A similar conclusion was drawn by
\inlinecite{Cheng2011} from the analysis of a different event.

\subsection{Shock Wave}
 \label{S-e1_shock}

The initial speed of the disturbance above the active-region core
must be equal to the fast-mode speed ($V_\mathrm{fast}$), to which
the estimated value of $\approx 1500$~km~s$^{-1}$ seems to
correspond. When the wave front left the region of high
$V_\mathrm{fast}$, such a high-speed wave must become strongly
nonlinear and could rapidly steepen into a shock of a moderate
intensity \cite{Afanasyev2013}. If the shock discontinuity was
formed before the passage of the wave through the arcade, then its
speed should change abruptly (the blue dashed curves in Figures
\ref{F-20110224_kinematics}a--\ref{F-20110224_kinematics}c);
otherwise, the kinematical curves should be smoother. As mentioned,
we cannot distinguish between the two options from the observations.

\begin{figure} 
  \centerline{\includegraphics[width=0.7\textwidth]
   {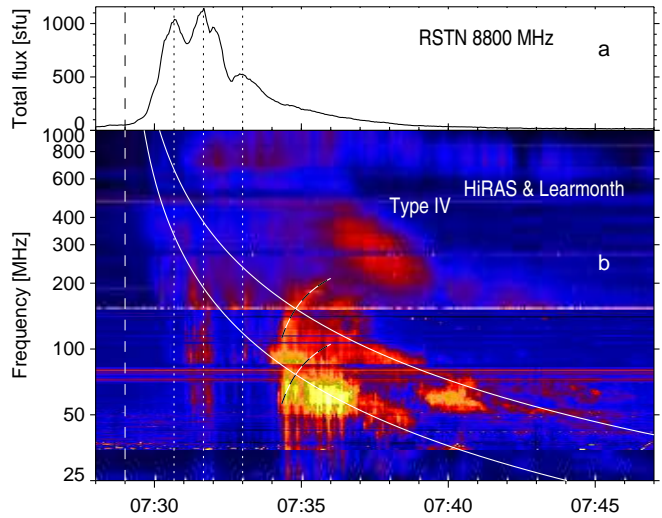}
  }
  \caption{The microwave time profile (a) and the dynamic spectrum
of the type II burst (b). The white curves outline the calculated
trajectory of the type II burst with a wave onset time
$t_0=\,$07:29:00$\,\pm20$~s marked with the dashed vertical line
and a density falloff exponent $\delta = 2.7$. The dashed curves
outline the reverse-drift portion and its expected fundamental
counterpart. The dotted vertical lines mark the microwave peaks to
compare them with type III bursts.}
  \label{F-20110224_spectrum}
  \end{figure}

On the other hand, all of the measured components of the non-radial
eruption reached the speeds of, at least, $V_{\mathrm{POS}} \approx
700$~km~s$^{-1}$ in the field of view of AIA ($|\mathbf{V}_{\max}|
\gsim 750$~km~s$^{-1}$). Such a speed of an ejecta above quiet-Sun
regions, where $V_\mathrm{fast}$ is considerably lower, is
sufficient to produce a bow shock. The presence of a shock is
confirmed by the appearance of a type II burst
(Figure~\ref{F-20110224_spectrum}b). To understand the history of
the shock wave, we consider its properties suggested by the type II
burst and a propagating `EUV wave'.

The dynamic spectrum in Figure~\ref{F-20110224_spectrum}b was
composed from the HiRAS and Learmonth data. For comparison,
Figure~\ref{F-20110224_spectrum}a presents the microwave time
profile at 8.8~GHz. The spectrum is complex. It shows a series of
type~III bursts, of which first three groups correspond to the
microwave peaks. Two branches of a type~IV emission are present: a
quasi-stationary type~IV burst around 800~MHz, and a drifting
type~IV, whose emission was strongest at 400--200~MHz (red). Such
drifting type~IVs are considered as manifestations of developing
CMEs.

\begin{figure} 
  \centerline{\includegraphics[width=0.6\textwidth]
   {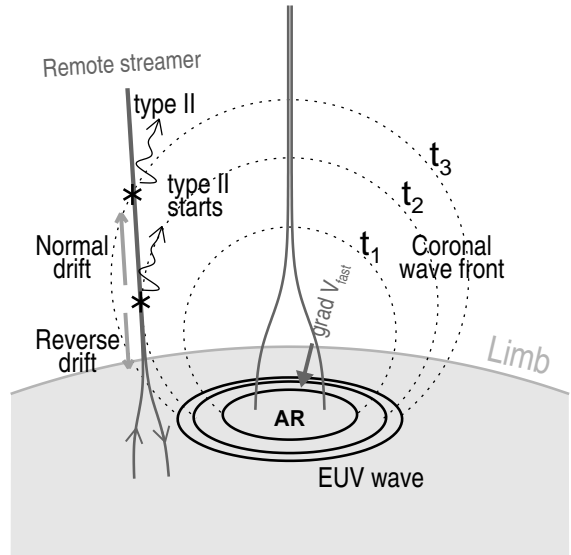}
  }
  \caption{Fast MHD shock wave excited by an impulsive eruption in an active
region (AR) and a type II burst produced by the shock in a remote
streamer. The expanding wave front is represented by the dotted
ovals at three consecutive times $t_1$, $t_2$, and $t_3$, and its
near-surface skirt (`EUV wave') is represented by the solid
ellipses.}
  \label{F-wave_cartoon}
  \end{figure}

A rather strong type~II burst, overlapping with an intense series
of type~IIIs, sharply started at 07:34 and became clearer after
07:37. We invoke the idea of \inlinecite{Uralova1994} that a type
II emission originates in a coronal streamer, being caused by a
shock front compressing its current sheet, thus producing a
flare-like process running along the streamer. A particularity of
the burst in Figure~\ref{F-20110224_spectrum}b is a feature with a
reverse drift from 100 to 200 MHz visible during 07:34--07:36
(dashed curve). A parallel dashed curve at 50--100 MHz outlines
its possible fundamental-emission counterpart, which is difficult
to detect in the figure. The reversely drifting feature looks
mirrored relative to the normal drift. \citeauthor{Grechnev2011_I}
(\citeyear{Grechnev2011_I,Grechnev2014_II}) interpreted such a
C-shaped feature as the onset of a type~II burst owing to the
collision of a quasi-perpendicular shock with a remote streamer.
The contact site of the shock front with the streamer bifurcates
and moves into opposite directions (Figure~\ref{F-wave_cartoon}).
While propagating into a higher-density medium, a shock wave
strongly dampens. Therefore, such reversely-drifting features are
usually marginal. The remarkably long lifetime of the reverse
drift here suggests that the shock was rather strong when the
type~II burst started.

We have outlined the trajectory of the type~II burst, using the
power-law fit. The pair of the white curves outlines the overall
evolution of the drift rate. The estimated wave onset time
$t_0=\,$07:29:00 is shown in Figure~\ref{F-20110224_kinematics}
with the vertical dotted line; the shaded interval presents the
uncertainty of $\pm 20$~s. This onset time corresponds to the rise
phase of the acceleration observed for the most dynamic leading
segment of the eruptive filament. Such a situation have been
observed in several different events, which we analyzed previously
(\citeauthor{Grechnev2011_I}, \citeyear{Grechnev2011_I,
Grechnev2014_I, Grechnev2014_II, Grechnev2013_20061213}), and
supports the impulsive-piston shock excitation scenario. The
outermost arcade loop, which we measured (blue in
Figure~\ref{F-20110224_kinematics}b), had a speed of $<
100$~km~s$^{-1}$ at that time; thus, the excitation of a bow shock
by the outer surface of the developing CME is excluded. The shock
excitation by the flare pressure pulse is unlikely, as shown in
Section~\ref{S-e1_eruption}.

These facts can be summarized in the following scenario. The
violent filament eruption with an acceleration of up to
$50g_\odot$ excited at about 07:29 a substantial MHD disturbance,
which initially resembled a blast wave. The wave steepened into a
rather strong shock before 07:34:30. This situation corresponds to
the impulsive-piston shock excitation scenario
(\citeauthor{Grechnev2011_I},
\citeyear{Grechnev2011_I,Grechnev2011_III};
\opencite{Afanasyev2013}).

\begin{figure} 
  \centerline{\includegraphics[width=\textwidth]
   {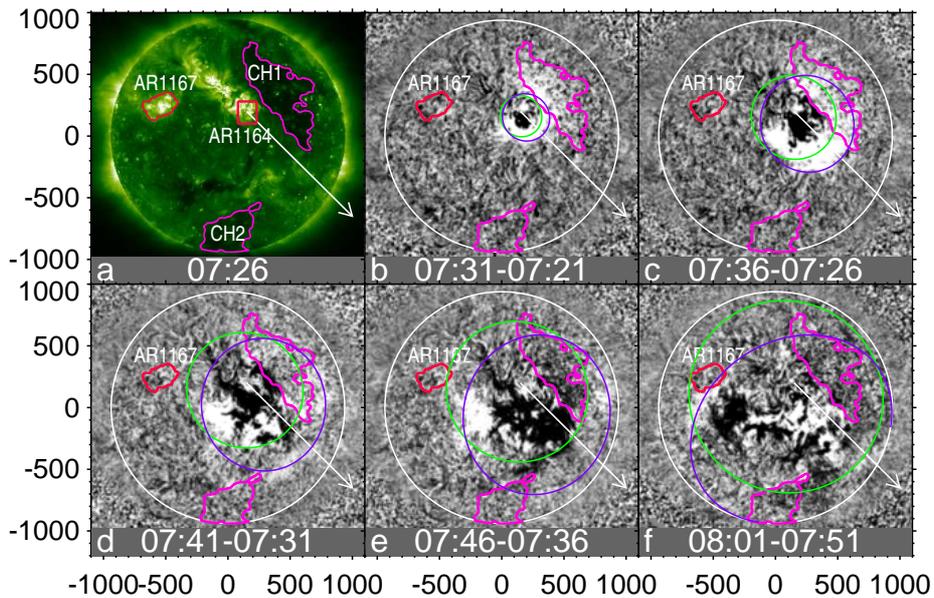}
  }
  \caption{STEREO-B/EUVI 195~\AA\ images of Event~I. The arrows
indicate the direction of the eruption. The contours outline
coronal holes (pink) and remote AR1167 (red). (a)~Pre-event image.
The red frame shows the field of view in
Figure~\ref{F-20110224_initiation}. CH1 and CH2 are the coronal
holes. (b)--(f)~`EUV wave' in running-difference image ratios. The
green ellipses outline an expected propagation of a blast shock
excited at the eruption site. The blue ellipses correspond to the
moving wave epicenter. The axes show the coordinates in arcsec
from the solar disk center.}
  \label{F-20110224_wave}
  \end{figure}

The global propagation of the disturbance is evidenced by the EUVI
195~\AA\ images in Figure~\ref{F-20110224_wave}. The pre-event
situation is shown in Figure~\ref{F-20110224_wave}a. The filament
eruption occurred in the region denoted with the red frame and
produced a disturbance, whose properties are typical of
shock-associated `EUV waves'. Its propagation was omnidirectional,
but not isotropic. The disturbance entered the adjacent coronal
hole, CH1, where it ran faster and had a lower brightness. These
properties are consistent with its MHD-wave nature. A higher
$V_\mathrm{fast}$ in a coronal hole relative to quiet-Sun regions
determines i)~a faster wave propagation and ii)~a lower Mach
number, \textit{i.e.,} a weaker plasma compression responsible for
the lower brightness (\textit{cf.} \opencite{Grechnev2011_III}).
The brightest portion of the `EUV wave' moved initially southwest
(Figures \ref{F-20110224_wave}b and \ref{F-20110224_wave}c). Then
the brightening propagating east became best visible (Figures
\ref{F-20110224_wave}d and \ref{F-20110224_wave}e). When the front
reached the remote AR1167 in Figure~\ref{F-20110224_wave}f, the
`EUV wave' apparently bypassed it (this property of `EUV waves',
also consistent with a non-uniform $V_\mathrm{fast}$ distribution,
as stated by \inlinecite{Thompson1999}). No brightening was
pronounced northeast of the eruption region; nevertheless, the
disturbance propagating in this direction can be followed from the
development of two lanes of dimming along AR~11164 in Figures
\ref{F-20110224_wave}c and \ref{F-20110224_wave}d. Long-lived
remote dimming can appear due to the pass of a shock wave indeed
\cite{Grechnev2013_20061213}. The running-difference images in the
right panel of the 20110224\_euvi\_195\_fulldisk.mpg movie show
complex disturbances in a large area southwest from the solar disk
center. The disturbances seem to move in different directions,
indicating reflection phenomena. This also supports the MHD-wave
nature of the `EUV wave'.

We tried to outline the global propagation of the `EUV wave' using
the power-law approximation, $r \propto t^{2/(5-\delta)}$ ($t$
time, $r$ distance, $\delta$ formal density falloff exponent)
expected for a blast wave, as we did previously. The result of
this attempt, with the same wave onset time $t_0=\,$07:29:00 and
$\delta = 2.0$, is shown with the green ellipses. Trying to follow
the northeastward wave signatures, such as the developing dimming
in Figures \ref{F-20110224_wave}b--\ref{F-20110224_wave}e and the
bright east portion of the `EUV wave', we miss its fastest
southwest part in Figures \ref{F-20110224_wave}c and
\ref{F-20110224_wave}d.

To reduce the mismatch, we displace the wave epicenter with a
speed of 250--300 km~s$^{-1}$ following the eruptive
filament---the blue ellipses in Figures
\ref{F-20110224_wave}b--\ref{F-20110224_wave}f. They match the
southern half of the `EUV wave' front considerably better.
However, the green ellipses better reproduce the wave propagation
northwest. The estimated surface propagation speed in Figures
\ref{F-20110224_wave}b--\ref{F-20110224_wave}d decrease from
$\approx 900$ to $\approx 650$~km~s$^{-1}$ along the arrow, and
from $\approx 560$ to $\approx 330$~km~s$^{-1}$ in the opposite
direction. All of them exceeded the expected $V_\mathrm{fast}$
above the quiet Sun. Let us see what the revealed properties of
the global disturbance mean.

The wave excited at $\approx $~07:29:00 steepened into the shock
before the onset of the type II burst at 07:34:30. This time is
between Figures \ref{F-20110224_wave}b and \ref{F-20110224_wave}c,
in which the `EUV wave' looks similar; probably, the shock
appeared still earlier, before the image at 07:31 in
Figure~\ref{F-20110224_wave}b. The global disturbance was also
observed on the Earth-facing side of the Sun by AIA in the 193 and
211~\AA\ channels (see the movies 20110224\_AIA\_211\_fulldisk.mpg
and 20110224\_AIA\_193\_fulldisk.mpg).

Previously we observed a slow progressive displacement of the
epicenter of a shock-associated `EUV wave' toward a region of a
higher $V_\mathrm{fast}$ (\citeauthor{Grechnev2011_I},
\citeyear{Grechnev2011_I, Grechnev2011_III,
Grechnev2013_20061213}). Such a displacement is due to refraction,
being an expected property of MHD waves. However, the wave
epicenter displaced in Figure~\ref{F-20110224_wave} considerably
faster and followed the motion of the eruption rather than the
$V_\mathrm{fast}$ distribution. While the ellipses with a moving
center matched the whole lower skirts of the wave fronts in
previous events, here a superposition of the two expanding fronts
seems to be present; the green ellipses correspond better to the
propagation northeast, and the blue ellipses are closer to the
southwest portions. Here, the whole wave front, whose shape was
initially close to a spheroid with a nearly radial axis, stretched
out afterwards, following the eruption. We have probably met, for
the first time, a situation, when the impulsively excited
blast-wave-like shock approached the bow-shock regime afterwards
(the rear shock propagated freely). A large tilt of the southern
wave front portion to the solar surface in Figures
\ref{F-20110224_sdo_images}g and \ref{F-20110224_sdo_images}h also
supports the bow-shock regime.

\subsection{CME}
 \label{S-e1_cme}

\begin{figure} 
  \centerline{\includegraphics[width=\textwidth]
   {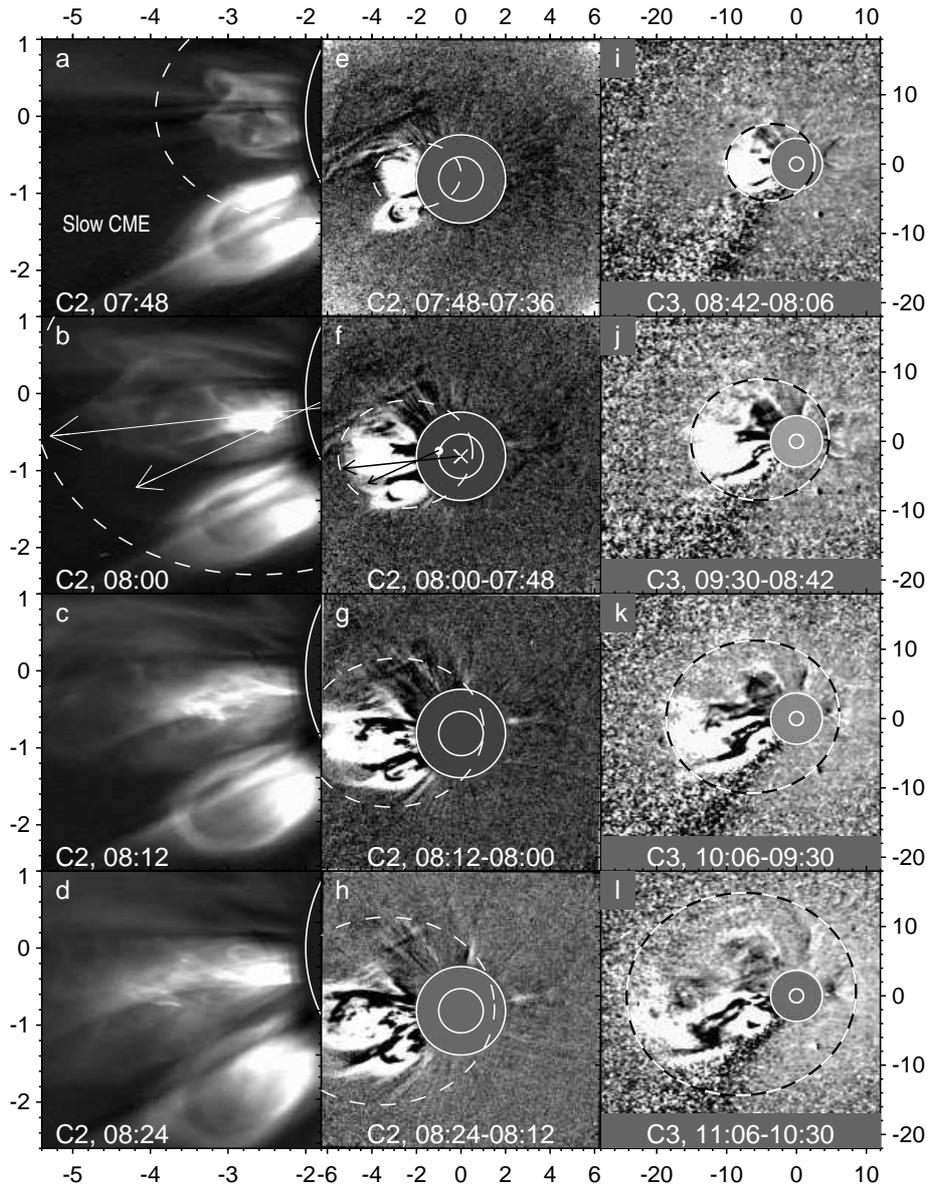}
  }
  \caption{The fast coronal transient in the LASCO/C2 (a--d and e--h) and
C3 (i--l) images. The structure of the fast CME and its collision
with the slow CME visible in C2 images (a--d) with a reduced field
of view. The dashed ovals outline the wave signatures in C2 (e--h)
and C3 (i--l) images shown with a full field of view as well as in
panels (a) and (b). The arrows in panels (b) and (f) indicate the
direction of the CME before the collision (shorter arrows) and
after it (longer arrows). The slated cross and the small filled
circle in panel (f) denote the solar disk center and the eruption
center, respectively. The smaller white open circles indicate the
location of the solar limb. The larger white open circles denote
the inner boundary of the field of view of the coronagraphs. The
axes show the coordinates in solar radii from the solar disk
center.}
  \label{F-20110224_lasco}
  \end{figure}

The eruption has produced a fast CME (the linear-fit speed of
1186~km~s$^{-1}$ according to the on-line CME catalog
(\opencite{Yashiro2004}; \opencite{Gopal2009_catalog};
\url{http://cdaw.gsfc.nasa.gov/CME_list/})) visible in the LASCO/C2
and C3 images presented in Figure~\ref{F-20110224_lasco}. There was
an additional complication due to a collision of the fast CME with
another, slow CME. The collision is clearly visible in the movies
available in the CME catalog.

The slow CME gradually ascended before the collision with a speed
of $\approx 115$~km~s$^{-1}$ (from 06:00 to 08:00), presumably
from below the streamer belt. The middle of this CME is crossed in
Figure~\ref{F-20110224_lasco}a, in projection, by a coronal ray,
located either in front of the slow CME or behind it. The leading
portion of the fast CME appeared in the field of view of LASCO/C2.
The dashed arc (same as in Figure~\ref{F-20110224_lasco}e)
outlines the expected position of the shock front, which we
discuss later. The expanding fast CME compressed the slow CME in
Figure~\ref{F-20110224_lasco}b. In the next images in Figures
\ref{F-20110224_lasco}c and \ref{F-20110224_lasco}d, the whole
slow CME was deflected.

The collision, which seems to be close to elastic one, has turned
the velocity vector of the fast CME. The change of the angle in
the LASCO's plane of the sky is shown by the arrows in Figures
\ref{F-20110224_lasco}b and \ref{F-20110224_lasco}f. The shorter
arrows correspond to the initial angle of the eruption of $\approx
25^{\circ}$ with respect to the east direction. The longer arrows
correspond to the final orientation of the fast CME of $6^{\circ}$
(\textit{i.e.}, the position angle of $96^{\circ}$ listed in the
CME catalog).

Figures \ref{F-20110224_lasco}a--\ref{F-20110224_lasco}d show the
structure of the fast CME. Its leading part appears to consist of
the expanding arcade loops. The tangled intertwisted structure of a
brighter core seems to be the relaxing flux rope formed from the
eruptive prominence. No cavity is pronounced.

The running differences in Figures
\ref{F-20110224_lasco}e--\ref{F-20110224_lasco}l reveal the traces
of the shock wave, mainly from the deflections of the coronal
rays. The fast CME in question was preceded by another one, and
therefore the outermost disturbances in Figures
\ref{F-20110224_lasco}e--\ref{F-20110224_lasco}g are irrelevant
(the southern part of the transient hints at the relevant traces).
We formally fit the kinematics of the leading edge of the
transient using the power-law fit with the same wave onset time,
$t_0 =$~07:29:00, as previously. The density falloff exponent,
$\delta = 2.70$ (corresponding to the mid-latitude Saito model
(\opencite{Saito1970}; see \opencite{Grechnev2011_I})), and the
reference distance were chosen to achieve a best fit of the
leading edge in all of the images in
Figure~\ref{F-20110224_lasco}.

The ovals outlining the whole shock front in
Figure~\ref{F-20110224_lasco} were computed from the fit of the
SOHO/LASCO images for the leading edge, and from the fit of the
STEREO-B/COR2 images for the lateral expansion. The ovals acceptably
match probable shock signatures, but it is difficult to distinguish
between the shock front and the trailing CME body in Figures
\ref{F-20110224_lasco}i--\ref{F-20110224_lasco}l (LASCO/C3). The
shock and CME had similar kinematics that corresponds to the
bow-shock regime. Eventually, the bow shock produced a glancing blow
on STEREO-B spacecraft on 26 February at 08:20, with ICME being
pointed by $\approx 20^{\circ}$ to the Sun--STEREO-B line, according
to our estimate.

\subsection{Summary of the Wave History in Event I}
 \label{S-summary_e1}

A violent prominence eruption above AR~11164 produced a strong
omnidirectional wave disturbance propagating with an initial speed
of $\approx 1500$~km~s$^{-1}$. When its front left the region of a
high $V_\mathrm{fast}$, the wave presumably steepened into a shock
and started to considerably decelerate in the directions, where it
was not followed by the eruption. Its signatures observed in
various spectral ranges propagated according to a power-law
kinematics expected for a shock wave. All of them had the same
onset time, corresponding to the peak acceleration of the
prominence. These signatures were i)~the trajectory of the type~II
burst, while its structure was consistent with shock encounters
with coronal streamers, ii)~the rear part of the `EUV wave',
iii)~the leading envelope of the white-light CME (which was, most
likely, super-Alfv{\' e}nic). The listed facts confirm that these
signatures were different manifestations of a single shock wave,
and that the prominence eruption was responsible for its
excitation.

Along with an omnidirectional power-law expansion with the same
onset time, the wave dome additionally expanded, following the
CME. This behavior was dissimilar to the events, which we analyzed
previously. This circumstance indicates that in this event, we are
probably dealing, for the first time, with a rapid transformation
of a leading part of an impulsively excited blast-wave-like shock
into a bow shock. Eventually, the shock was detected at the Earth
orbit.

\section{Event II: 11 May 2011}
 \label{S-e2}

To verify and elaborate the conclusions drawn from the preceding
event, which was rather strong, now we consider the weak 11 May
2011 event (GOES importance of  B8.1). In this event no HXR
emission was detected. It was associated with the eruption of a
filament centered at N25\,W54, between small ARs 11207 and 11204.
The eruption produced a CME. We analyze the observations of this
event carried out by SDO/AIA, SOHO/LASCO, NoRH, and SSRT.

We have produced the NoRH 17 GHz images in steps of 60~s with an
integration time of 10~s to enhance the sensitivity. These images
had a displaced center enabling us to track the erupting filament
as long as possible. The SSRT images at 5.7 GHz covered the main
phase of the flare with intervals from 3 to 8 min. The advanced
technique used to produce the SSRT images and to calibrate both
the SSRT and NoRH images is described by
\inlinecite{Kochanov2013}.

\subsection{Filament Eruption}

\begin{figure} 
  \centerline{\includegraphics[width=\textwidth]
   {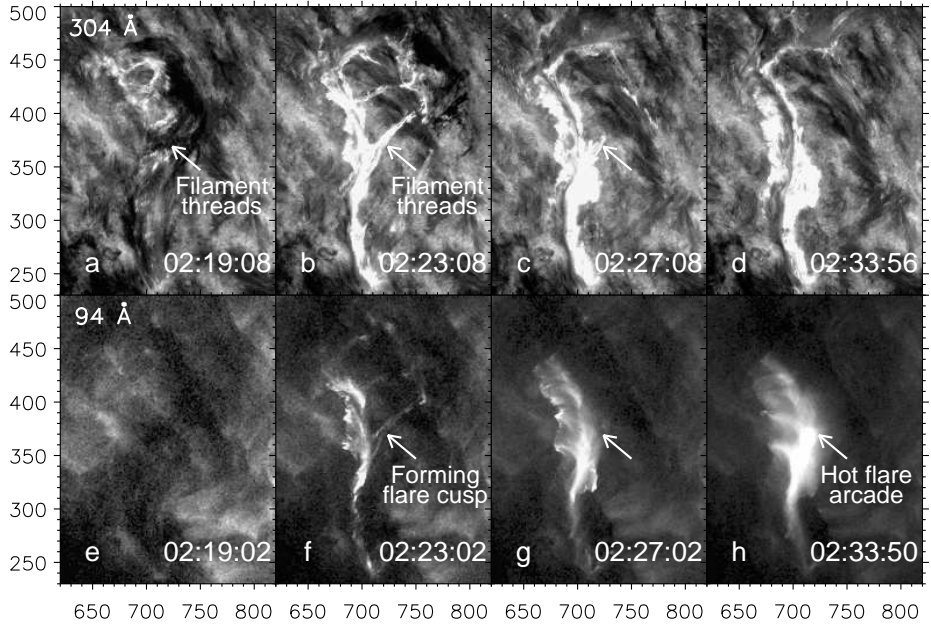}
  }
  \caption{Filament eruption on 11 May 2011 observed by SDO/AIA
in 304~\AA\ (top row) and in 94~\AA\ (bottom row). The positions
of all the arrows are the same.}
  \label{F-20110511_304_94}
  \end{figure}

Figure~\ref{F-20110511_304_94} presents the initial stage of the
eruption observed in the 304~\AA\ channel (characteristic
temperature $5 \times 10^4$~K, top row) and in the 94~\AA\ channel
(6.3~MK, bottom row). The rising filament body is dark in 304~\AA\
and blocks radiation from its background indicating its
temperature of $\lesssim 10^4$~K. A similar appearance of the
stretching filament threads in 304 and 94~\AA\ suggests a wide
temperature range in their brightened parts. Their mutual
correspondence indicates the presence of $\sim 6$~MK plasma. The
configuration of the stretched filament threads resembles a flare
cusp. The bundles of the threads descending from the cusp are
rooted in different ribbons, whose development is shown in the
upper row. The lower row presents the development of the flare
arcade above the ribbons. The later 94~\AA\ images in Figures
\ref{F-20110511_304_94}g and \ref{F-20110511_304_94}h confirm that
the hot flare arcade developed at the same place, where the scusp
was present. Thus, the flare arcade started to form from the
threads belonging to the filament body. The standard-model
reconnection between the legs of the pre-flare arcade, which
embraced the filament from above, presumably started later on.

\begin{figure} 
  \centerline{\includegraphics[width=\textwidth]
   {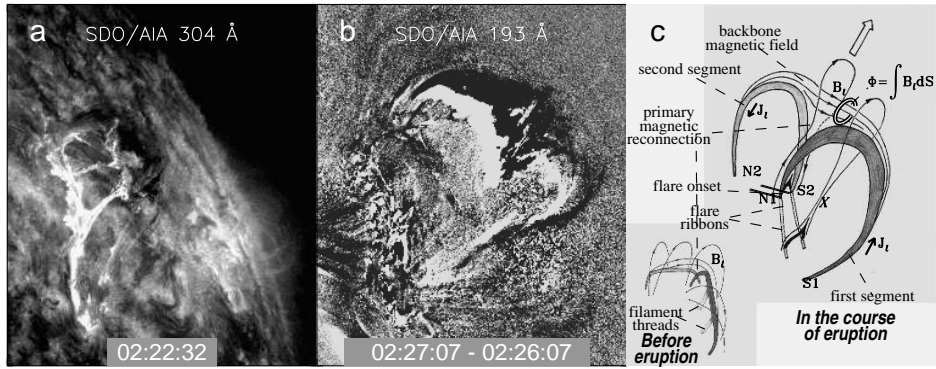}
  }
  \caption{The 11 May 2011 event. Comparison of the eruptive filament
observed with SDO/AIA in the 304~\AA\ channel (a) and a later
running-difference 193~\AA\ image (b) with the dual-filament model
(c; adapted from Uralov \textit{et al.} (2002)).}
  \label{F-dual_filament}
  \end{figure}

Figure~\ref{F-dual_filament} compares the actually observed
eruption with the dual-filament CME initiation model
(Section~\ref{S-challenges}; \opencite{Uralov2002}).
Figure~\ref{F-dual_filament}a shows the filament in 304~\AA\
(similar to Figure~\ref{F-20110511_304_94}b). A running-difference
193~\AA\ image in Figure~\ref{F-dual_filament}b reveals two
filament segments, a slower thicker north one and a faster thinner
south one. The heated south threads discussed above seem to be
shared by both segments. A striking similarity between the
observations and model supports the formation of the flare cusp
from the filament threads. The model predicts strengthening the
propelling force, when two segments combine.

\subsection{Expansion of the Filament and the Overlying Arcade}

\begin{figure} 
  \centerline{\includegraphics[height=0.82\textheight]
   {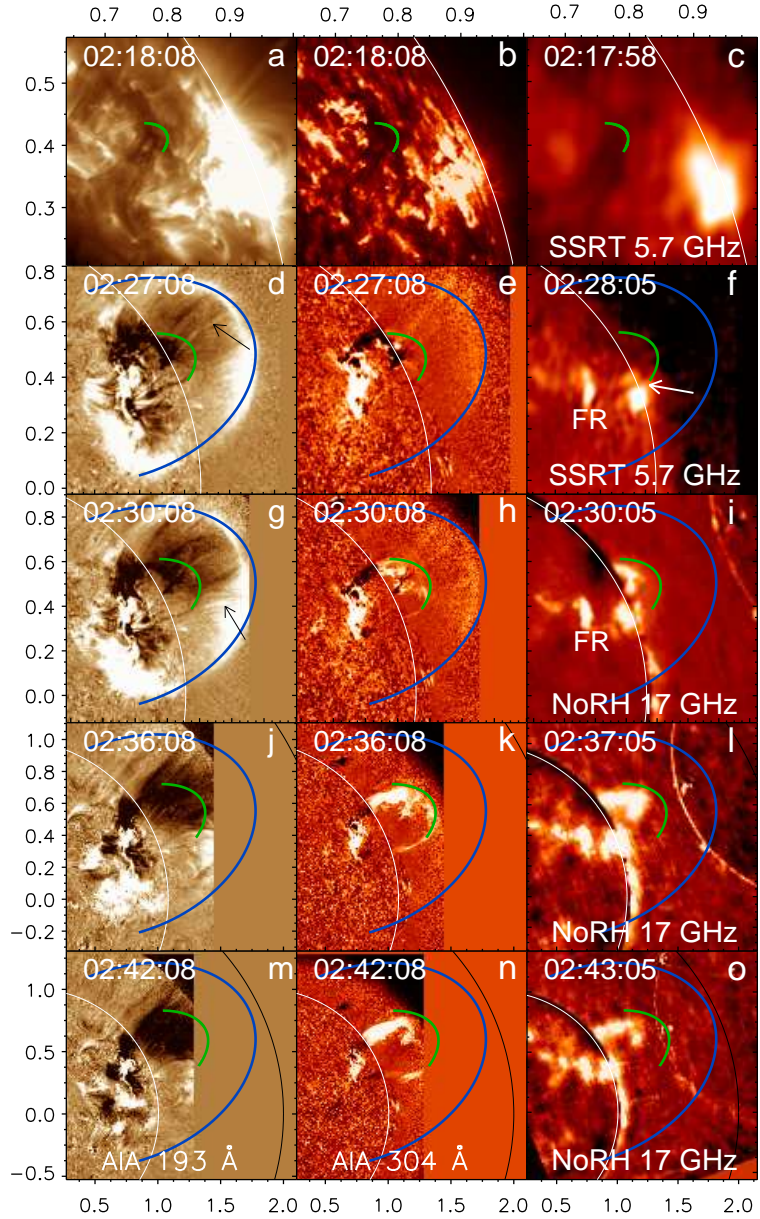}
  }
  \caption{Expansion of the erupting filament and the arcade above
it in SDO/AIA 193 and 304~\AA\ as well as microwave images: SSRT
at 5.7 GHz (c,f) and NoRH at 17 GHz (j,l,o). The AIA images in the
second to bottom rows are the ratios to the images observed in the
corresponding channels at 02:18:08. All of them were rotated to
02:27:00. The quiet-Sun's disk was subtracted from the microwave
images. The flare region is labeled `FR' in panels (f) and (i).
All the images are resized according to the measured kinematics in
Figure~\ref{F-20110511_kinematics} to keep the visible size of the
expanding eruption fixed. The blue arc outlines the arcade. The
green arc outlines the filament. The white arc outlines the solar
limb. The black arc in two lower rows corresponds to the inner
boundary of the LASCO/C2 field of view of $2R_{\odot}$. The axes
show the coordinates in solar radii from the solar disk center.}
  \label{F-sdo_norh}
  \end{figure}

The expansion of the eruptive structures is presented in
Figure~\ref{F-sdo_norh}. The eruptive filament is clearly visible in
304~\AA\ and somewhat poorer in 193~\AA. The thick north filament
segment looks very similar in the 17 GHz and 304~\AA\ images. We
measured from these images the kinematics of the expanding
filament's top in the direction between the two segments with an
uncertainty $< 8$~Mm. We also made the measurements of a poorer
accuracy from the 17 GHz images, whose field of view reaches $1.5
R_{\odot}$ (422~Mm from the eruption site), until 02:43:05.

\begin{figure} 
  \centerline{\includegraphics[width=0.6\textwidth]
   {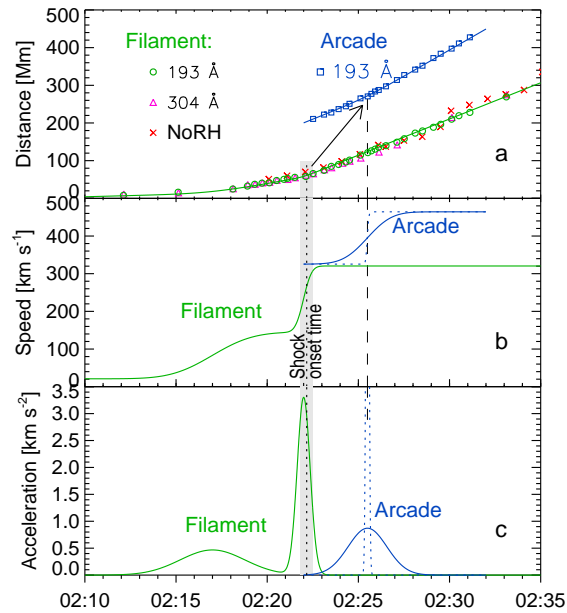}
  }
  \caption{Measurements of the kinematics for the eruptive
filament and arcade in Event~II from the SDO/AIA images in 304 and
193~\AA\ and the NoRH 17~GHz images. The symbols show the initial
straightforward measurements. The solid curves present the fit.
The thin dashed lines correspond to a possible shorter
acceleration of the arcade. The dotted vertical line denotes the
estimated shock onset time, and the shaded interval presents its
uncertainty. The dashed vertical line corresponds to the
acceleration peak time of the arcade. The arrow in panel (a)
indicates a disturbance propagating from the erupting filament to
the arcade.}
  \label{F-20110511_kinematics}
  \end{figure}

Figure~\ref{F-20110511_kinematics} shows the kinematic plots of the
measured filament's part and a faint oval above it associated with
the top of a pre-eruption arcade, as discussed below. The solid
curves present the fit. Initially, the filament gradually rose with
a speed of $\approx 21$~km~s$^{-1}$, which is typical of the
initiation stage. The filament underwent an acceleration of
0.5~km~s$^{-2}$ around 02:17:00, and sharply ($\approx 3.3$
km~s$^{-2}$) accelerated again at 02:22:05 $\pm 5$~s to a final
velocity of $V_{\max} \approx 320$~s$^{-1}$, which did not change
after 02:25. The spine of the coupled filament segments expanded
slightly faster than the clearly visible measured feature, with a
velocity of $340$~s$^{-1}$, which we use in further measurements.

The expansion of the arcade was measured in a nearly the same
direction as that of the filament. Initially, the oval on top of
the arcade was not sharp, and the uncertainties were larger, about
15~Mm. According to the fit in Figure~\ref{F-20110511_kinematics},
the arcade expansion accelerated from $325 \pm 5$ km~s$^{-1}$ to
$465 \pm 5$ km~s$^{-1}$. The observations allow us to estimate
only the upper limit for the duration of the acceleration and the
lower limit for its maximum of $\gsim 0.9$ km~s$^{-2}$. Similarly
to the situation in Event~I, the use of a considerably stronger
acceleration does not result in a pronounced difference in the
outline of the arcade in the images. The center time of the
acceleration is certain, 02:25:30~$\pm 10$~s.

The sequence of phenomena presented in
Figure~\ref{F-20110511_kinematics} is similar to Event~I
(\textit{cf.} Figure~\ref{F-20110224_kinematics}). The eruptive
filament in Event~II underwent a strongest acceleration around
02:22:05 and produced a wave disturbance with an onset time $t_0
=$~02:22:10~$\pm 20$~s (the vertical dotted line). The disturbance
indicated by the arrow in Figure~\ref{F-20110511_kinematics}a
traveled about 200~Mm with an average plane-of-the-sky speed of
$\approx 1000$~km~s$^{-1}$, reached at 02:25:30 the arcade (the
vertical dashed line), which gradually expanded above the
filament, and impulsively accelerated its lift-off.

Figure~\ref{F-sdo_norh} shows the expansion of the eruptive filament
and the arcade above it as observed by SDO/AIA in 193~\AA\ (left
column), in 304~\AA\ (middle column), and in microwaves by SSRT and
NoRH (right column). All of the images are progressively resized
according to the measured kinematics, to keep the size of the
eruption unchanged. The top of the expanding filament and the arcade
are outlined with the oval arcs, whose radii were calculated from
the corresponding fit in Figure~\ref{F-20110511_kinematics} with a
mentioned correction.

Microwave images in the right column show the erupting filament
from its initial position up to large distances. The dark filament
is visible in the SSRT images at 5.7 GHz with a higher contrast
against the solar disk, whose brightness temperature is
$T_{\mathrm{QS}(5.7)} = 16000$~K, than in the NoRH images with
$T_{\mathrm{QS}(17)} = 10000$~K. The thin south filament segment
is detectable in the 5.7 GHz image in Figure~\ref{F-sdo_norh}f due
to the overlap with a bright region on the limb. Then we use the
higher-sensitivity NoRH images at 17 GHz, where the bright
filament is better visible against the sky, than in the SSRT ones.
The thick north filament segment appears similar in the 17 GHz and
304~\AA\ images. The measured kinematics of the eruptive filament
shows a good correspondence with all the images. Both filament
segments in 304~\AA\ were joined by an oval spine corresponding to
the backbone field in the dual-filament model in
Figure~\ref{F-dual_filament}c.

The faint bright oval in Figure \ref{F-sdo_norh}d and
\ref{F-sdo_norh}g is similar to the arcade top in Event~I. Here, the
orientation of the flux rope's axis indicated by the filament and
flare ribbons was close to the plane of the sky. Thus, the oval
surrounding the eruptive filament cannot be the cross section of a
flux rope. On the other hand, nearly radial bright thin structures
faintly visible inside the oval (\textit{e.g.}, those indicated by
the arrows in Figures \ref{F-sdo_norh}d and \ref{F-sdo_norh}g)
resemble the outer arcade loops, whose planes had small angles with
the line of sight. This identification corresponds to the
orientation of the filament and ribbons. A separatrix surface should
exist above the arcade. Thus, the arcade top prevented any plasma
motions from outside into its interior. Its expansion resulted in
the appearance of a compression region constituted by the swept-up
plasmas on its top.

The expansion of the arcade `membrane' results in a rarefaction in
the volume enclosed by the arcade. As the 193~\AA\ ratio images in
the left column show, dimming behind the expanding arcade
developed. The observations shed light on its cause. The
brightness $B$ in EUV images is proportional to the column
emission measure. The brightness in the center of an expanding
volume of a linear size $L$ with a fixed total number of emitting
particles should be $B \propto EM/A \propto n^2L = (N_0/V)^2L
\propto 1/L^5$. The expansion alone results in this dramatic
brightness decrease. The resulting density depletion should cause
a siphon effect to fill the dimmed volume by plasma flowing from
below. This circumstance is consistent with plasma flows from
dimmings revealed by \inlinecite{HarraSterling2001}. This
phenomenon seems to be a secondary effect of the expansion due to
the CME lift-off and is expected to be commonly present.

The same factor of $B \propto 1/L^5$ applies to the microwave
brightness. For this reason, eruptions observed in EUV and
microwaves rapidly fade away. Note that the Thomson-scattered
light responsible for the white-light transients is controlled by
a much softer factor of $B \propto 1/L^2$, which grants the
opportunities to observe CMEs up to very large distances from the
Sun.

As mentioned, the arcade accelerated at 02:25:30~$\pm 10$~s to a
final speed of $465 \pm 5$ km~s$^{-1}$ with a measured
acceleration of $\gsim 0.9$ km~s$^{-2}$. Its actual duration could
be less, and the maximum value could be higher. In any case, the
acceleration of the arcade lagged behind that the prominence, as
in Event~I. The temporal relation between the observed phenomena
in Figure~\ref{F-20110511_kinematics} and the estimated velocities
of the erupting filament, arcade, and wave demonstrate that the
wave drove the arcade, but not \textit{vice versa}. The prominence
was more dynamic, which indicates that it was the major driver of
the whole eruptive event. The fast prominence eruption with an
acceleration up to 3.3 km~s$^{-2}$ ($\approx 12g_{\odot}$) could
produce a shock wave.

\subsection{Shock Wave}

\begin{figure} 
  \centerline{\includegraphics[height=0.8\textheight]
   {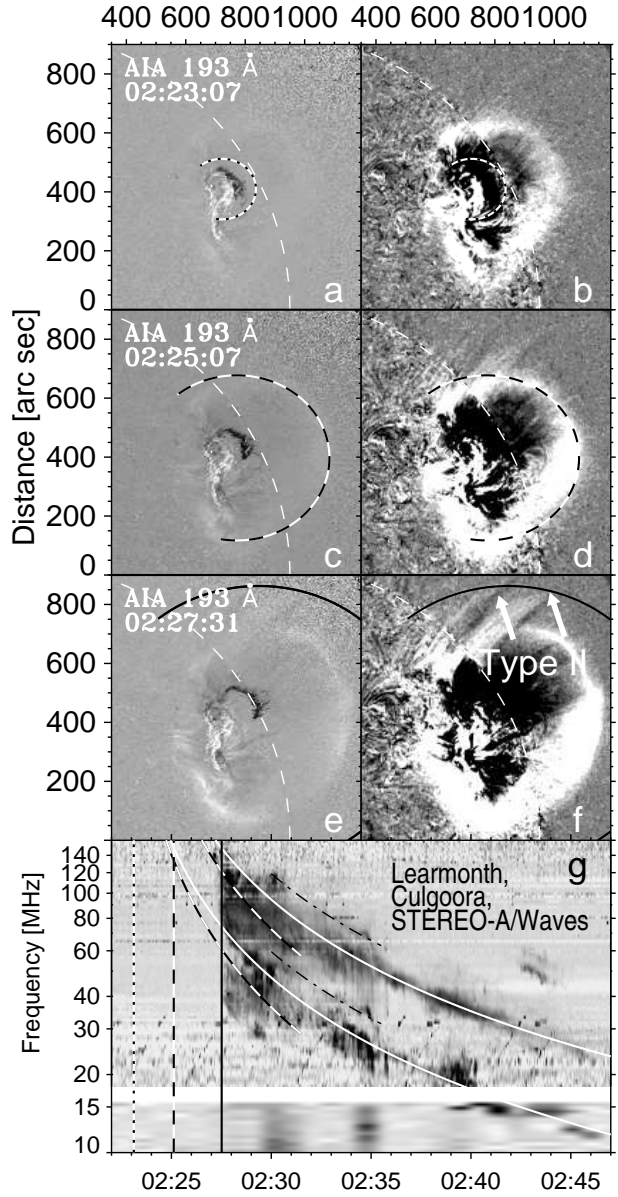}
  }
  \caption{Expanding wave front in the SDO/AIA 193~\AA\ images in
comparison with a type II burst. The left column shows fixed-base
ratios at 193~\AA\ to reveal the eruptive filament. The right
column shows running-difference ratios in 193~\AA\ to reveal the
expanding arcade and deflected small streamers north to it. The
thick ovals present the calculated wave fronts with $t_0 =
\,$02:22:10, $\delta = 2.7$. The thin dashed arc denotes the solar
limb. (g)~Dynamic spectrum of the type II burst. Three pairs of
the curves with the same shock onset time, $t_0$, outline
different harmonic pairs of the bands presumably emitted from
different streamers hit by the same shock front. The vertical
lines mark the times of the images above. The colors and line
styles of the ovals correspond to those of the vertical markers on
the spectrum.}
  \label{F-aia_wave_spectrum}
  \end{figure}

There are manifestations of a shock wave excited in this event
indeed. Figure~\ref{F-aia_wave_spectrum}g shows a dynamic spectrum
composed from the records made with the Learmonth, Culgoora, and
the STEREO-A/WAVES \textit{Radio and Plasma Wave Investigation}
\cite{Bougeret2008} spectrographs. At 02:27:30, a type II burst
suddenly started. It had a few pairs of bands, of which three are
outlined with the curves of different line styles. Two pairs of
the bands began nearly simultaneously and resembled
band-splitting, which is usually related to the emissions from
electrons accelerated in the upstream and downstream regions of
the shock \cite{Smerd1974}. However, the distance between the
bands belonging to different pairs was large. The traditional
interpretation accounts for two pairs of bands only.
Alternatively, \inlinecite{Grechnev2011_I} proposed an explanation
of a large band-splitting by a passage of the shock front over
remote streamers located close to each other (the geometry similar
to that shown in Figure~\ref{F-wave_cartoon}). This process can
account for a more complex multi-band structure of a type II
burst. Figure~\ref{F-aia_wave_spectrum}f and the movies
20110511\_AIA\_193\_eruption.mpg and
20110511\_AIA\_193\_spectrum.mpg really show two small ray-like
structures located aside of the eruption. They were inflected by a
disturbance propagating from the erupting structure. The contours
calculated for a single decelerating blast wave match the
evolution of the drift rate for all bands of the type II burst. We
have estimated the shock onset time $t_0 =\,$02:22:10~$\pm 20$~s,
which corresponds to the strongest acceleration of the erupting
filament and to various shock manifestations.

The calculated propagation of the shock front is shown in
Figure~\ref{F-aia_wave_spectrum} and movie
20110511\_AIA\_193\_spectrum.mpg in comparison with the dynamic
spectrum. The panels of each row present the same AIA 193~\AA\
image processed in different ways. The images in the left column
are fixed-base ratios, which show the erupting filament better.
The images in the right column are running-difference ratios,
which allow one to see the expanding arcade and an `EUV wave'
developing at its base. The ovals represent the calculated
portions of the wave fronts with $t_0 = \,$02:22:10 and $\delta =
2.7$. Here we do not analyze the near-surface propagation of the
shock wave, and therefore the ovals are not closed. The three
pairs of images present different times: shortly after the wave
onset (Figures \ref{F-aia_wave_spectrum}a and
\ref{F-aia_wave_spectrum}b), during the passage of the wave
through the arcade (Figures \ref{F-aia_wave_spectrum}c and
\ref{F-aia_wave_spectrum}d), and at the onset of the type II burst
(Figures \ref{F-aia_wave_spectrum}e and
\ref{F-aia_wave_spectrum}f). These times are marked on the dynamic
radio spectrum in Figure~\ref{F-aia_wave_spectrum}g with vertical
lines, whose styles and colors correspond to the ovals. The arrows
in the third row indicate the inflected small streamers, in which
the sources of the type II burst could be located. The estimates
of the distances and velocities from dynamic spectra are uncertain
by an unknown density multiplier; the points in question are the
wave onset time and the trajectory of the type II burst.

Figure~\ref{F-aia_wave_spectrum} confirms that the sharply erupting
filament excited a shock wave, which passed through the arcade,
accelerated its expansion, and then reached two small remote
streamers, thus causing the generation of the type II burst in them.
Two additional facts support the presence of a shock wave. i)~A
slower surface trail (`EUV wave') of the expanding bright oval in
the right column corresponds to the idea of \inlinecite{Uchida1968},
initially proposed for Moreton waves. ii)~The
20110511\_AIA\_193\_eruption.mpg movie shows that the southeast
portion of the `EUV wave' is reflected from a coronal hole at about
02:48, and a backward-reflected front runs slower than the incident
front (\textit{cf}. \opencite{Gopalswamy2009_I}) that is expected
for a shock wave \cite{Grechnev2011_I}. These facts confirm the MHD
shock-wave nature of the propagating `EUV wave'.

\begin{figure} 
  \centerline{\includegraphics[width=\textwidth]
   {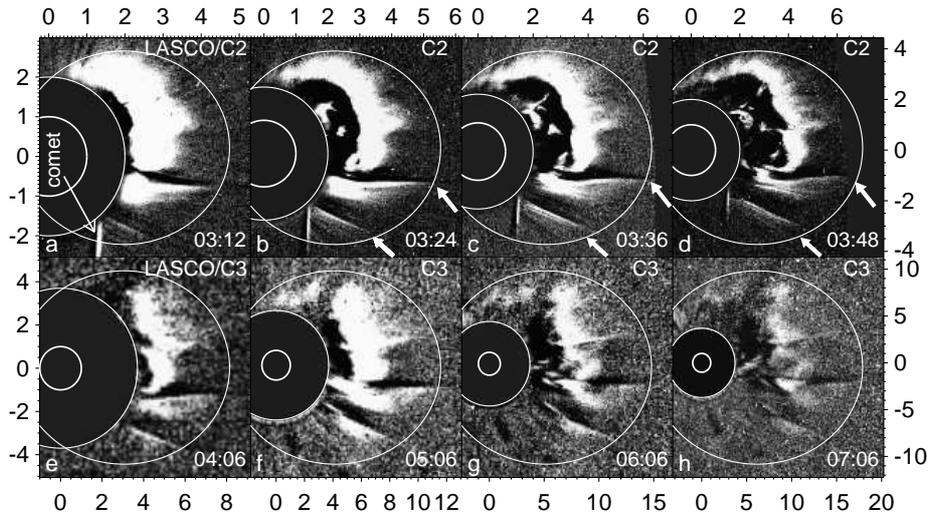}
  }
  \caption{Traces of the expanding wave front in running-difference
ratios produced from LASCO/C2 images (top row) and C3 ones (bottom
row). The ovals represent the calculated wave fronts. All the images
are resized to keep the visible size of the expanding wave front
fixed. The axes show the coordinates in solar radii from the solar
disk center. The white solid circles denote the solar limb and the
inner boundaries of the LASCO/C2 and C3 field of view.}
  \label{F-lasco_wave}
  \end{figure}

We have applied a power-law fit with the same parameters as in
Figure~\ref{F-aia_wave_spectrum} to the LASCO observations in
Figure~\ref{F-lasco_wave}. To see the expanding features better
and facilitate their comparison with each other, all of the images
are progressively resized to keep the size of the expanding wave
front unchanged. The ovals represent the calculated fronts of the
shock wave, which decelerated from the initial velocity of $>
1200$~km~s$^{-1}$ to $\approx 550$~km~s$^{-1}$ in the latest
images, becaming comparable to the solar wind speed. The wave is
manifested at large distances in the deflections of coronal rays
indicated by the arrows in the upper row (\textit{cf}.
\opencite{Sheeley2000}; \opencite{Vourlidas2003};
\opencite{Gopalswamy2009_II}). The ovals encompass the outer
boundary of the CME toward the regions above the north pole and
match the wave traces ahead of the CME up to $17R_{\odot}$. The
shock wave in this event eventually decayed into a weak
disturbance.

\subsection{CME}

\begin{figure} 
  \centerline{\includegraphics[width=\textwidth]
   {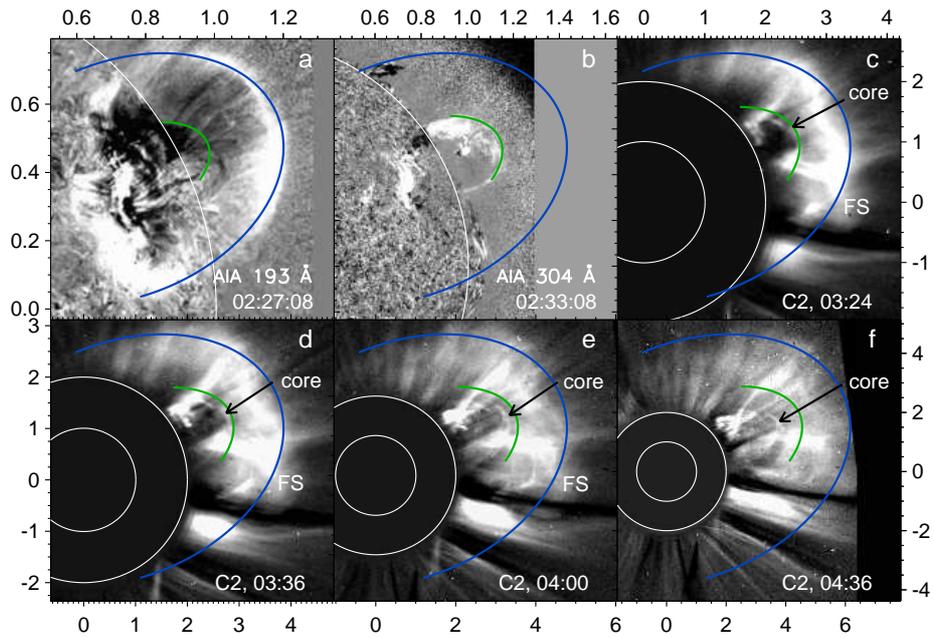}
  }
  \caption{CME observed by LASCO/C2 in the 11 May 2011 event (c--f)
in comparison with the arcade (a, AIA 193~\AA) and erupting
filament (b, AIA 304~\AA). All of the images are resized to keep
the size of the CME fixed. The ovals outline the expanding frontal
structure FS (blue) and core (green) according to the measured
speeds of the arcade and filament, respectively. The axes show the
distances from the solar disk center in solar radii. The white
solid circles denote the solar limb and the inner boundary of the
LASCO/C2 field of view.}
  \label{F-cme}
  \end{figure}

The resizing representation can help to find the nature of the
components of the CME produced in this event. Figure~\ref{F-cme}
presents fixed-base image ratios to compare the expanding arcade
in Figure~\ref{F-cme}a and the coupled filament in
Figure~\ref{F-cme}b with the CME structure in Figures
\ref{F-cme}c--\ref{F-cme}f. The ovals outlining the arcade (blue)
and filament (green) help in identifying the CME components with
their progenitors. The expansion speed used to resize the images
and to plot the ovals was constant, 340~km~s$^{-1}$ for the
filament and 465~km~s$^{-1}$ for the arcade.

The ovals outlining the filament and the arcade match the CME core
and frontal structure (FS) in earlier LASCO images, respectively.
The frontal structure consisted of nearly radial loops, whose planes
had acute angles with the line of sight. These facts confirm our
identification of the bright oval in Figure~\ref{F-cme}a with an
arcade, and show that the arcade was a progenitor of the frontal
structure. The core originated from the eruptive filament, as
commonly recognized. The features leading the frontal structure,
like the distorted streamers indicated by the arrows, reveal the
wave ahead of the CME.

The core lags behind the dashed outline in later images, Figures
\ref{F-cme}c--\ref{F-cme}f. The filament, which initially drove the
expansion of the arcade, relaxed and decelerated. The arcade became
the inertial frontal structure. The impulsive-piston-driven shock
ran well ahead of the CME, like a decelerating blast wave.

\subsection{Summary of the Wave History in Event II}
 \label{S-summary_e2}

Although Event~II (B8.1) was considerably weaker than Event~I
(M3.5), the observations reveal the same wave excitation scenario
in both events. Here, the erupting filament produced a wave
disturbance with an onset time $t_0 =$~02:22:10. The propagating
wave reached the arcade, which gradually expanded above the
filament, and accelerated its lift-off. Then the wave deflected
two remote streamers. The wave propagation visible in EUV, as well
as its farther signatures in white-light LASCO images, follow a
single power-law distance-time plot expected for a shock wave,
with the same $t_0$. The trajectory of the type~II burst also
corresponds to a power-law shock-wave kinematics with the same
$t_0$, and its structure corresponds to the shock encounter with
the mentioned streamers. Thus, all of these signatures were
different manifestations of a single shock front, which originated
in the filament eruption.

In Event~II, the CME body was slow (465~km~s$^{-1}$, as we showed),
and the shock wave eventually decayed into a weak disturbance. The
different speeds of the CME bodies determined the different wave
evolutions in the two events, although their origins were similar.

\section{Discussion}
 \label{S-discussion}

The chain of phenomena observed in Events I and II suggest the
following scenario. The activation of a filament in the initiation
phase precedes the flare, being probably related to heating in the
filament or/and its coronal environment. These processes, which
respond in a gradual rise of the SXR emission (Event~I), commence
the final formation of an eruptive flux rope. The initial rise of
the filament stretches magnetic threads passing through its body
and ending at the solar surface, so that they cross each other and
form a cusp-like configuration (Events I and II) above the future
flare site. The cusp is constituted by quasi-antiparallel magnetic
fields, which start to reconnect. This results in heating in this
region (Event~II). A violent MHD instability (probably, the torus
one, Event~I) develops, increasing the acceleration of the
filament. The flare develops, being slightly delayed after the
filament acceleration (Event~I). The sharply accelerating filament
i)~forces the magnetic structures above it to expand, and
ii)~produces a substantial MHD disturbance, which rapidly steepens
into a blast-wave-like shock. Then the acceleration of the
filament ceases, its magnetic structure relaxes, and its
combination with the structures expanding above it constitute a
CME. Eventually, the shock ahead of the CME either changes to the
bow-shock regime, if the CME is fast (Event~I), or decays
otherwise (Event~II). Its flanks and rear propagate freely. This
picture of a flare-related eruption seems to be different from
non-active-region eruptions of quiescent filaments.

\subsection{Formation of a Flux Rope and Flare Initiation}

A low-lying progenitor of a flux rope often shows up as a
pre-eruption filament (prominence). It is frequently considered as
a passive part of a larger flux rope, whose expansion creates a
CME. By contrast, a pre-eruption filament in an AR carries large
electric currents, as the linear force-free approach ($\alpha =
\mathrm{const}$) confirms. The density of the electric current is
proportional to the magnetic field strength, being typically
larger near the solar surface. The development of an MHD
current-driven instability (kink or torus) is only possible, if
the distribution of the $\alpha$ parameter is inhomogeneous. Even
in the situation of the non-linear force-free magnetic field, the
major current-carrying part of a flux rope is expected in its
bottom part, \textit{i.e.}, in a low-lying structure like a
pre-eruption filament, where $|\alpha|$ is maximum. Thus, a
filament (or its analog) in an active region appears to be the
major progenitor of the eruptive flux rope, whose instability
increment is governed by stronger, lower, smaller-scale magnetic
fields, while the influence of the weaker, larger-scale
environment is less important.

The magnetic field and plasma continuously occupy the whole magnetic
environment of an AR. They are topologically bounded by the
separatrix surface, which confines the coronal cavity surrounding a
prominence, and probably develops later into the frontal structure
of a CME. The cavity is similar to a magnetic `cocoon' enveloping
the pre-eruptive filament and consisting of sheared current-carrying
loops, which are rooted in the photosphere. The cavity is
traditionally identified with a `perfect' magnetic flux rope, which
is only rooted in the photosphere by two ends, being disconnected
elsewhere.

A real pre-eruption filament has numerous lateral connections to
the photosphere with its threads (barbs) on both sides of the
magnetic neutral line. The perfect flux rope not yet exists; its
progenitor is the magnetic structure of the filament inside the
magnetic cocoon. Their combination is the initial structure in the
CSHKP model, in which magnetic loops of the expanding cocoon
stretch out, reconnect under the eruptive filament, and form the
expanding flux rope. The eruptive filament is located inside the
cocoon; its external boundary is the separatrix surface. The joint
expansion of all of these structures is observed as a CME.

In our consideration, the formation of the perfect flux rope is
necessary to trigger the main stage of the `standard' flare. The
flux rope starts to form from breaking its lateral connections to
the photosphere leading to the formation of primary flare ribbons
by reconnection between the filament threads crossing each other,
rather than CSHKP-reconnection of magnetic field lines, which
initially were not shared by the filament. The opposite-polarity
magnetic field of the filament threads, which participate in the
primary reconnection, pass through the body of the filament and
embrace it, being its intrinsic components. The primary
reconnection is not a simple reconnection of a single pair of flux
tubes, as in the tether cutting model. Instead, this is a wavelike
series of reconnection events between the threads, running along
the filament and the polarity inversion line.

In non-active-region eruptions of quiescent filaments, which are
easier to observe, the increments of an instability
(\textit{e.g.}, a torus one) corresponding to the filament and
cocoon should not be much different from each other. The filament
and the larger cocoon expand jointly as a single, self-similar CME
structure practically from the very start. The radial profile of
the plasma velocity inside the CME is linear relative to the
expansion center. These circumstances produce an impression of a
passive role of the filament in the eruptive process.

\subsection{Development of Shock Waves and Accompanying Phenomena}

Eruptions of large quiescent filaments might produce shocks far from
the Sun, but unlikely in the low corona. The situation is different,
if a CME forms in an AR, where the eruptive filament initially moves
considerably earlier and faster, than the structures above it. The
expansion of the developing flare-related CME is not self-similar at
this stage.

\begin{figure} 
  \centerline{\includegraphics[width=0.6\textwidth]
   {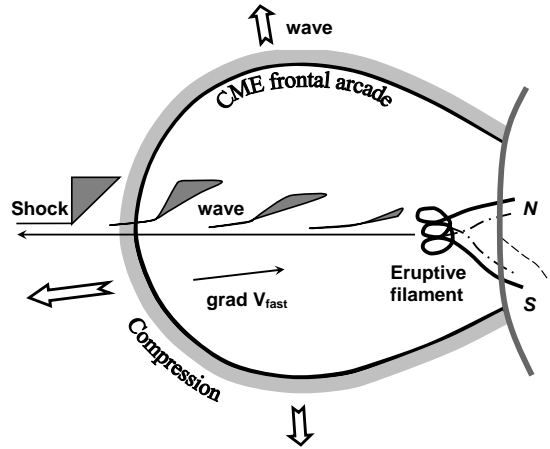}
  }
  \caption{Presumable scenario of the wave excitation and its
steepening into a shock. The expanding separatrix surface,
associated with the CME frontal arcade, confines the eruptive
filament. Passing through the arcade, the shock front accelerates
its expansion.}
  \label{F-shock_scenario}
  \end{figure}

The intensity of a disturbance produced by an impulsively erupting
flux rope, that develops from the filament, is proportional to the
squared accelerations, with which its major and minor radii
increase. The sharpest portion in the velocity profile of the
disturbance propagating away from such a piston forms
approximately at the same time as its acceleration reaches its
maximum. A discontinuity starts to form at this place.
$V_\mathrm{fast}$ is high above an AR and steeply falls off both
upward and laterally. While propagating in such medium, the
leading wave packets of the spheroidal MHD disturbance decelerate,
being rapidly overtaken by trailing ones. This process produces a
`jam' effect, which distorts the disturbance profile. The wave
becomes strongly nonlinear (see Figure~\ref{F-shock_scenario}),
and the discontinuity of a moderate intensity is formed in $\sim
10^2$~s \cite{Afanasyev2013}. The expansion of the arcade above
the erupting flux rope is initially passive and occurs due to the
deformation of magnetic fields surrounding the developing flux
rope. When passing through the arcade, the shock front somewhat
accelerates its expansion and runs afterwards ahead of the
developing CME like a decelerating blast wave.

The expanding CME bubble bounded by the outer separatrix surface
extrudes surrounding magnetoplasmas almost omnidirectionally, thus
forming an extensive compression zone around it. The front of this
zone is a weak discontinuity expanding with the ambient
$V_\mathrm{fast}$. Within this zone $V_\mathrm{fast}$ is higher
than the ambient one before the event. The growing layer of the
compressed plasma on top of the CME bubble makes its boundary
visible. The compressed oval layers might be a rather common
phenomenon, disclosing developing CMEs (see, \textit{e.g.},
\opencite{Cheng2011}). Such bright ovals could occasionally be
observed by SOHO/EIT as sharp `S-waves', although the probability
to catch them with EIT was low (7\% according to
\opencite{Biesecker2002}) due to its insufficient imaging rate. As
an `S-wave' can actually be the compressed plasma layer swept-up
by a gradually expanding arcade, relating `S-waves' to shock waves
(\textit{e.g.,} \opencite{Ma2011}) might be not correct. The
brightenings associated with the near-surface parts of the
separatrix surfaces, or `quasi-stationary EIT waves', appear to
represent the `field-stretching' effects
\cite{ChenFangShibata2005}. An example is a brightening south from
the eruption site in Event~II, around $[750^{\prime \prime},
80^{\prime \prime}]$ from the solar disk center. It is visible in
the left panel of the 20110511\_AIA\_193\_eruption.mpg movie.

The traveling `EUVI waves' in Events I and II were, most likely,
near-surface skirts of the expanding domes of MHD shock waves, in
agreement with \inlinecite{Uchida1968} and
\inlinecite{Thompson1999}. Their MHD-wave nature is confirmed by
their velocities, reflection phenomena, and accordance of their
propagation up to very large distances (more than the solar
hemisphere in Event~I) with the fast-mode speed distribution.
These facts is difficult to reconcile with non-wave
interpretations, such as the mentioned field-stretching model of
\inlinecite{ChenFangShibata2005} or the interchange-reconnection
model of \inlinecite{Attrill2007}. While alternative
interpretations of `EUV waves' (see, \textit{e.g.},
\opencite{WillsDavey2009}; \opencite{Gallagher2011} for a review)
are mostly focused on near-surface phenomena, we established in
Events I and II the kinematic correspondence between them, the
trajectories and structures of type II bursts, and the CME
expansion.

The shock front ahead of the CME gains energy from the trailing
`piston', which spends it to sweep up plasmas. Thus, the shock
front kinematically resembles a self-similar blast wave
propagating in plasma with a density falloff $\delta \approx 2.7$,
close to the Saito model. For the shock propagation along the
solar surface, $\delta = 0$ might be expected, but this is not the
case due to the radial inhomogeneity of the solar corona (see
\opencite{Grechnev2008shocks}, Section 4.5). The kinematics of the
shock front, which is not followed by a separatrix surface
laterally, can formally be described with $\delta > 0$. The
results of this approach are close to the modeled shock
propagation \cite{AfanasyevUralov2011, Grechnev2011_III}.

The rapidly decelerating near-surface portion of the wave front can
be observed in the H$\alpha$ line as a fast Moreton wave, usually
not far from the eruption site, and as a slower `EIT wave' in EUV at
larger distances. Having encountered a coronal streamer, the shock
front compresses its current sheet causing magnetic reconnection
there. The cumulation effect intensifies the flare-like process
running along the streamer. A type II burst appears (see
Figure~\ref{F-wave_cartoon}).

The bow shock continuously driven by the outer surface of a
super-Alfv{\'e}nic CME (see, \textit{e.g}.,
\opencite{VrsnakCliver2008}; \opencite{Reames2009}) can be
actualized in the case of a supersonic plasma flow around the
outer surface of the CME bubble. Two conditions are important for
this regime: i)~the existence of a stagnation point in the plasma
flow on the surface of the body, and ii)~the velocity of the
stagnation point must exceed the ambient $V_\mathrm{fast}$. Such a
plasma flow does not yet exist at the early CME formation stage.
In the bow-shock regime, the CME size and speed ($V_{\mathrm{CME}}
> V_{\mathrm{fast}}$) determine the position and intensity of the
stationary shock ahead of the CME. The bow-shock problem does not
consider the kinematical differences of the structural CME
components preceding the appearance of the shock, including the
large difference of their accelerations from the self-similar
regime. With strong accelerations measured for the impulsively
erupting filaments in Events I and II, the impulsive-piston
mechanism excited shock waves effectively and rapidly. The
kinematics of the whole CME determines whether or not the frontal
portion of the wave transforms into the bow shock afterwards.

The CME frontal structure in Event~II had a nearly constant speed of
465~km~s$^{-1}$ up to $7R_{\odot}$, unlikely exceeding the Alfv{\'
e}n speed (see, \textit{e.g.}, \opencite{Mann2003}). The shock wave
quasi-freely propagated well ahead of the CME, being detectable up
to $17R_{\odot}$ in Figure~\ref{F-lasco_wave}, where it decelerated
to $\approx 530$~km~s$^{-1}$, comparable to the solar wind speed.
This shock wave eventually decayed into a weak disturbance. By
contrast, the blast-wave-like shock impulsively excited in Event~I
later showed indications of its approach to the bow-shock regime,
even when propagating over the solar surface. The transition into a
bow shock at large distances from the Sun was anticipated by
\citeauthor{Grechnev2011_I} (\citeyear{Grechnev2011_I,
Grechnev2013_20061213}). The early appearance of the bow shock in
Event~I was favored by the conspicuously non-radial motion of the
eruption and its high speed, which was rapidly reached at a small
height. As a result, the rapidly moving southwest part of the `EUV
wave' in later EUVI images of this event probably was a near-surface
trail of a bow shock.

The deceleration of coronal transients can be due to different
forces. They are the gravity and magnetic tension of structures,
which remain anchored on the Sun. The plasma extrusion by the
expanding CME and the aerodynamic drag from the solar wind flowing
around the CME (\textit{e.g.}, \opencite{Chen1996};
\opencite{VrsnakGopalswamy2002}) also spend its energy. We cannot
distinguish which retarding force dominated at each stage and to
detect the transition from one deceleration regime to another.
Presumably, the CME core decelerated in Event~II (Figures
\ref{F-cme}d--\ref{F-cme}f) mainly due to the magnetic tension;
deceleration due to the plasma extrusion dominated during the
initial extra-radial expansion of the CMEs; and the aerodynamic
drag was significant in the bow-shock regime.

One more consequence of the plasma extrusion by expanding CMEs is
the development of dimming. The brightness in EUV and SXR images
is proportional to the column emission measure, \textit{i.e.}, the
squared density. The CME expansion alone results in a dramatic
rarefaction of the involved volume observed as dimming. A strong
pressure gradient appears and causes a secondary plasma outflow in
the footprint regions of a CME \cite{HarraSterling2001}. The
utmost velocity of the outflow is limited by the sound speed
\cite{Jin2009}.

\section{Summary}
 \label{S-conclusion}

The observations considered in Sections \ref{S-e1} and \ref{S-e2}
appear to confirm the scenario of \inlinecite{Hirayama1974}
incorporated into the CSHKP model with further elaborations
mentioned in Section~\ref{S-introduction}. The results update the
model and specify the flare--CME relations, as well as the
excitation and evolution of associated shock waves.

The observations indicate that the flare arcade starts to form from
the threads of the filament body. This implies involvement in the
eruptive-flare reconnection processes of considerably lesser-height,
\textit{i.e.}, stronger magnetic fields with higher gradients. This
can help to understand some observational challenges.

The flux ropes were found by \inlinecite{Qiu2007} to be mainly
formed via reconnection process, and independent of pre-existing
filaments, whose role was unclear. Our results indicate that flux
ropes form in the same way from their filament-like progenitors,
while the possibility to observe a filament depends on its
predominant temperature and density and, possibly, other
conditions.

All observational facts considered here confirm that we dealt with
fast MHD shock waves, initially excited in the low corona by
sharply erupting flux ropes, and neither by a flare pressure pulse
nor by the outer surface of a CME. The initial impulsive-piston
shock excitation during the early flare rise was responsible, with
minor variations, for the shocks observed in a number of different
events, ranging from the GOES B class up to the X class
(\opencite{Meshalkina2009}; \citeauthor{Grechnev2011_I},
\citeyear{Grechnev2011_I, Grechnev2014_II,
Grechnev2013_20061213}). The concept, which related the source of
a type II emission to the current sheet of a coronal streamer
stressed by a shock front, has accounted for the structural
features of the observed type II bursts. The evolution of shock
waves and their manifestations in EUV and white-light images, and
dynamic radio spectra, have been quantitatively reconciled with
each other. They get further confirmation from a recent study of
\inlinecite{Kwon2013}.

\begin{acks}
We appreciate the efforts of the colleagues operating SSRT and
NoRH. We thank G.~Rudenko, S.~Anfinogentov, K.-L.~Klein,
K.~Shibasaki, L.~Kashapova, N.~Meshalkina, and Y.~Kubo for their
assistance and discussions, and anonymous reviewers for useful
remarks. Our special thanks to reviewer~2 for the valuable
recommendations, which significantly helped us to bring the paper
to its final form. We are grateful to the instrumental teams
operating SDO/AIA, STEREO/SECCHI, Wind/WAVES and S/WAVES, RHESSI,
SOHO/LASCO (ESA \& NASA), NICT, Culgoora Radio Spectrograph, USAF
RSTN network, and GOES satellites for the data used here. We thank
the team maintaining the CME Catalog at the CDAW Data Center by
NASA and the Catholic University of America in cooperation with
the Naval Research Laboratory. This study was supported by the
Russian Foundation of Basic Research under grants 11-02-00757,
12-02-00037, 12-02-33110-mol-a-ved, 12-02-31746-mol-a, and
14-02-00367; the Integration Project of RAS SD No.~4; the Program
of basic research of the RAS Presidium No.~22, and the Russian
Ministry of Education and Science under projects 8407 and
14.518.11.7047.

\end{acks}

\end{article}

\end{document}